\documentclass[letterpaper,usenatbib]{mn2e}
\usepackage{graphicx}
\usepackage{rotating}
\usepackage{aas_macros}
\pdfoutput=1
\def\etal{{\it et al. }} 
\title[Globular Clusters and the Spheroid of NGC~3923] 
{Gemini/GMOS Spectroscopy of the Spheroid and Globular Cluster System of NGC~3923} 

\author[Norris \etal] {Mark A. Norris$^{1}$\thanks{m.a.norris@dur.ac.uk}, Ray M. Sharples$^{1}$, 
Terry Bridges$^{2}$, Karl Gebhardt$^{3}$, Duncan A. Forbes$^{4}$,\newauthor
Robert Proctor$^{4}$, Favio Raul Faifer$^{5}$, Juan Carlos Forte$^{6}$,
Michael A. Beasley$^{7}$,\newauthor Stephen E. Zepf $^{8}$, David A. Hanes$^{2}$\\
\\
  $^1$ Department of Physics, University of Durham, 
  South Road, Durham DH1 3LE\\
  $^2$ Department of Physics, Queen's University, 
  Kingston, ON K7L 3N6, Canada\\
  $^3$ Astronomy Department, University of Texas, 
  Austin, TX 78712, USA\\
  $^4$ Centre for Astrophysics \& Supercomputing, Swinburne University, 
  Hawthorn, VIC 3122, Australia\\ 
  $^5$ IALP - CONICET, Argentina\\
  $^6$ Facultad de Cs. Astronomicas y Geofisicas, 
  UNLP, Paseo del Bosque 1900, La Plata, and CONICET, Argentina\\
  $^7$ Instituto de Astrofisica de Canarias, 
  La Laguna 38200, Tenerife, Spain\\ 
  $^8$ Department of Physics and Astronomy, Michigan State University, 
  East Lansing, MI 48824, USA\\
}

\begin{document}

\date{Accepted 2007 ***. Received 2007 ***; in original form ***}

\pagerange{\pageref{firstpage}--\pageref{lastpage}} \pubyear{2007}

\maketitle

\label{firstpage}

\begin{abstract}
We present a technique to extract ultra-deep diffuse-light
spectra from the standard multi-object spectroscopic observations used to 
investigate extragalactic globular cluster (GC) systems. This technique 
allows a clean extraction of the spectrum of the host 
galaxy diffuse light from the same slitlets as the GC targets. 
We show the utility of the method for investigating 
the kinematics and stellar populations of galaxies at 
radii much greater than usually probed in longslit 
studies, at no additional expense in terms of telescope 
time. To demonstrate this technique we present 
Gemini/GMOS spectroscopy of 29 GCs associated with the
elliptical galaxy NGC 3923. We compare the measured stellar 
population parameters of the GC system with those of the 
spheroid of NGC 3923 at the same projected radii, and find
the GCs to have old ages $>$
10 Gyr, [$\alpha$/Fe]$\sim$0.3 and a range of 
metallicities running from [Z/H] = -1.8 to +0.35.
The diffuse light of the galaxy is found to have 
ages, metallicities and [$\alpha$/Fe] abundance ratios 
indistinguishable from those of the red GCs.

\end{abstract}


\section{Introduction}
In recent years great strides have been made in 
the spectroscopic study of globular clusters (GCs), 
and we have begun to build up large samples of 
spectroscopically determined ages, metallicities 
and alpha-element abundances with the aim of using 
these data to infer star formation histories for their host 
galaxies \citep{Kissler-Patig98,Kuntschner02,
Puzia04,Puzia05,Strader05,Pierce3379,Pierce4649,
Cenarro07,Strader07}. GCs are particularly suitable 
for this analysis because they can be assumed to be 
simple stellar populations (SSP) where all of their stars 
formed at one epoch from a single cloud of gas with an 
almost uniform metallicity. This makes them much simpler 
to model than the integrated stellar populations of the 
diffuse light of galaxies, where many star formation 
events at different epochs greatly complicate analysis 
of the spectra.

The great promise of this approach to simplify the 
examination of the star formation history of galaxies 
is however dependent on one critical assumption: that 
the stellar populations of GCs are representative 
of the field stars that form the bulk of a galaxies' stellar
population. This would occur naturally if 
GCs form with some fixed proportion to those stars that 
form in unbound star forming regions during a star 
formation event and with similar element 
abundances. There are several observations that appear 
to bear out this assumption, including observations 
by \cite{Larsen2000} that show that the number of 
Young Massive Clusters (YMCs) correlates with the star 
formation rate per unit area in the galaxies that 
host them. These YMCs are believed to be the low 
redshift analogues to the star clusters that survived 
for a Hubble time to form the mostly old GC populations 
seen in galaxies today. The almost constant value 
observed for the number of GCs normalized to the total 
baryonic mass of a galaxy \citep{McLaughlin99} also 
provides strong support for the idea that GCs closely 
trace the major star formation events of a galaxy.

A more direct way to approach this issue is to examine the 
ensemble parameters of large numbers of GCs and compare 
them with those measured for the host galaxies' diffuse 
light. If the GCs do indeed act as a faithful tracer 
of the different star formation events in a galaxy's 
history, then the average GC stellar population
should closely resemble that of the overall galaxy.

To date little work has been done in comparing the stellar 
population parameters measured from line index analyses for 
GCs with those determined for the host galaxy diffuse stellar 
component. Principally this is because at the galactocentric
radii where the GCs selected for spectroscopic analysis are 
found, the surface brightness of the underlying galaxy halo 
light is very faint.
	In \citet{NSK06} the stellar populations along the major 
and minor axes of the isolated S0 galaxy NGC 3115 were 
compared to the stellar populations of the GC population of NGC 3115 
taken from \citet{Kuntschner02}. For this galaxy it was 
found that the stellar population along the minor axis 
at $\sim$2R$_{e}$, was indistinguishable from that of the 
more metal rich GCs in terms of age ($\sim$12 Gyr), [Z/H] 
($\sim$-0.5 dex) and [$\alpha$/Fe] ($\sim$0.3 dex). Furthermore,
 at larger radii along the minor axis the stellar population 
became more metal-poor, implying that if it were possible 
to observe the halo stellar population at even larger 
radii it may begin to resemble the more metal poor GCs.
These results hint at a strong connection between the
formation of the spheroid of NGC 3115 and its GC system.

Ideally we would like to repeat this type of analysis for 
many galaxies to see how universal this behaviour is. However 
the large additional burden of telescope time required for 
the longslit measurements on top of the long exposure times 
required for the MOS (Multi-Object Spectroscopy) GC 
investigations makes this approach impractical.
	In this paper we present a new technique 
that allows us to combine both parts of this type of 
study into one set of observations with no extra 
cost in terms of telescope time. As an additional benefit, this 
technique can also be used to extract kinematic parameters for the host 
galaxy at radii well beyond those commonly achieved for standard 
longslit studies.

For this study we examine the galaxy NGC 3923
(see Table~\ref{tab:ngc3923_detail} for details), which is a large nearby 
(D $\sim$ 21Mpc) elliptical galaxy with a prominent shell 
structure \citep{Malin83}. NGC 3923 is the brightest galaxy
of an average sized group with a low early-type fraction, perhaps 
suggesting that the group is dynamically young \citep{Brough06}.

An HST ACS study of the GC system of NGC 3923 by
\citet{sikkema06} confirmed previous observations  
\citep{zepf94,zepf95} that the GC system has a bimodal colour 
distribution and an unsually high specific frequency (S$_N$$>$5) for an 
early type galaxy in a lower density environment.
\cite{sikkema06} searched for
evidence of a GC sub-population formed during the merger event
thought to be responsible for the creation of the shell structures. 
They concluded that their photometric data did not support the presence 
of any younger GC population with ages similar to that of the shell 
structures, which they believe to have formed 0.8-1.2 Gyr ago. They did 
however find that the bimodal V-I colour distribution and radial density 
profiles of the blue and red GCs were typical for old GC systems in ellipticals.
\citet{Sikkema07} used the same HST ACS data to study the
shell structures of NGC 3923 finding them to be similar in colour or slightly
redder than the galaxy stellar light. 

Other studies have examined the stellar population and kinematics
of the central regions of NGC 3923. \citet{Thomas05} found the nuclear
regions to have an age of 3.3 Gyr, [Z/H] = +0.62 and [$\alpha$/Fe] = 0.31,
while \citet{Denicolo05} using a similar method found values of
2.6 Gyr, [Z/H]  $>$0.67 and [$\alpha$/Fe]  = 0.14 respectively. 

This paper is structured as follows. Section 2 provides details 
on the data used to investigate the GC and diffuse 
stellar content of NGC 3923; Section 3 describes the 
method used to extract extra information on the underlying 
galaxy spectrum from a given set of MOS observations; Section 4 
briefly describes kinematic results derived 
using this technique, and provides an example of 
the line index analysis that is possible using this method.
In Section 5 we discuss the implications of these results,
Section 6 provides some concluding remarks.


\begin{table}
 \centering
 \begin{minipage}{140mm}
  \caption{NGC 3923 Basic Parameters}
   \begin{tabular}{@{}lr@{}}
    \hline
     Parameter    & Value      \\
     \hline
     Right Ascension (J2000)                 & 11$^{h}$51$^{m}$01.8$^{s}$\\
     Declination (J2000)                     & -28$^{\circ}$48{'}22{"}\\
     $l$							& 287.28   \\
     $b$							& 32.22 \\
     Morphological Type                      & E4	\\
     Magnitude				     & 10.8 B mag\\
     Major Diameter                          & 5.9 arcmin\\
     Minor Diameter                          & 3.9 arcmin \\
     Heliocentric Radial Velocity            & 1739   $\pm$9 \,kms$^{-1}$\\
     J-Band Half-Light Radius    	     & 43.8" $\dagger$\\
     \hline
     Table data from NED: \\
     http://nedwww.ipac.caltech.edu/ \\
     Except:\\
     $\dagger$ 2 Micron All Sky Survey \\
     www.ipac.caltech.edu/2mass/ \\
    
      \label{tab:ngc3923_detail}
\end{tabular}
\end{minipage}
\end{table}


\section{Observations and Data Reduction}
Whilst we present a specific example using data obtained from
the Gemini-GMOS instrument, the approach described here
should be equally applicable to similar MOS intruments such
as FORS1/2 on the VLT or LRIS/DEIMOS on Keck.

All observations described here were
taken with the Gemini South Multi-Object-Spectrograph (GMOS)
\citep{Hook04} as part of Gemini program GS-2004A-Q-9.

Pre-imaging of NGC 3923 for object selection was undertaken
on 2004 January 19 and consisted of 4$\times$200s 
exposures in Sloan g$^\prime$ and 4$\times$100s in r$^\prime$ and i$^\prime$ for 
each of 3 fields (one central, one SW and one NE of the 
galaxy centre). For a thorough discussion of the 
procedure used to reduce the pre-imaging and select 
GC candidates see \cite{Forbes04} and \cite{Bridges06}.
A full examination of the photometric properties of the GC system
of NGC 3923 and other galaxies investigated in this project will 
be presented in a forthcoming paper Faifer et al. (in prep).

\begin{figure} 
   \centering
   \begin{turn}{0}
   \includegraphics[scale=0.5]{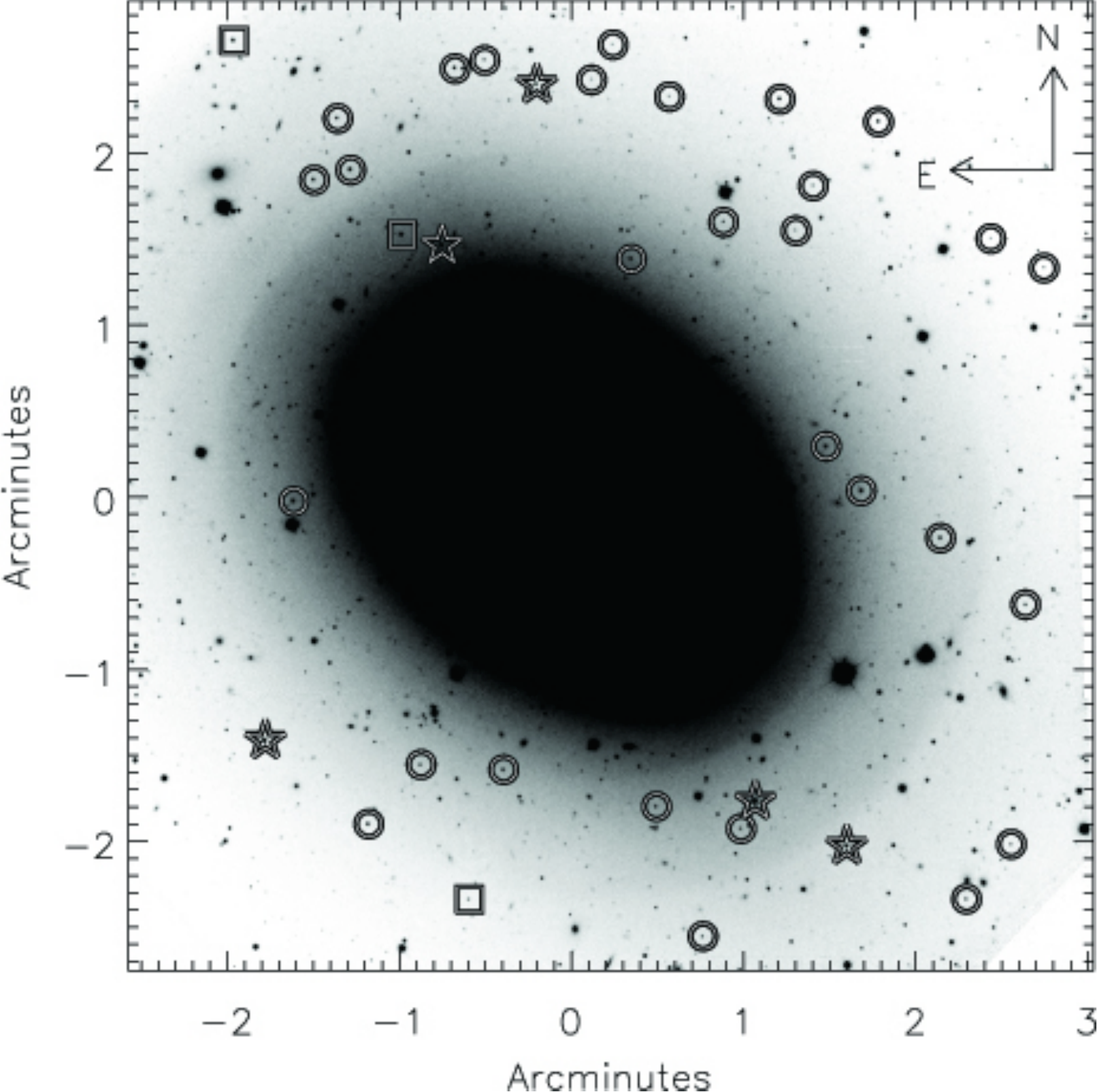}
   \end{turn}
    \caption{NGC 3923 GMOS i$^\prime$ band image. Circles show the 
    position of photometrically chosen targets later confirmed 
    by spectrocopy to be GCs, stars show targets found to be 
    Milky Way stars and squares are targets found to be 
    background galaxies/QSOs. The problem of placing a slitlet at large 
    enough galactocentric radii to produce an uncontaminated 
    sky spectrum when using modern MOS instruments with 
    fields of view around 6$\times$6 arcmin$^{2}$ is apparent. 
    Also visible are the shell structures of NGC 3923 on the major axis.}
   \label{fig:mos_field}
\end{figure}

GMOS Nod and Shuffle masks were produced for each of the 
three fields in addition to one MOS mask for the central
pointing (see Fig.~\ref{fig:mos_field}). Since it is the MOS
mask that provides the data  described here,
we shall leave any further discussion of the Nod and
Shuffle masks and the kinematic investigation of NGC 3923 
derived from them to a later paper (Norris et al. in prep.). 

The MOS mask
consisted of 37 slitlets of width 1 arcsec by a minimum 
length of 4 arcsecs. The mask was exposed using the
B600\_G5303 grism for 8$\times$1800s at a central 
wavelength of 500nm and 8$\times$1800s at a central 
wavelength of 505mn (to cover the CCD chip gaps), yielding 
8 hours of on-source integration. The seeing ranged from 0.6 - 0.9 
arcsec during the observations. Bias frames, flat fields and
copper-argon (CuAr) arc spectra were observed throughout
the observations as part of the Gemini baseline calibrations.
The MOS spectra produced typically
cover the wavelength range 3900-5500\,\AA\, although because
the wavelength range depends on slit position, some 
spectra start at $\sim$3500\,\AA\, while others end at 
$\sim$7200\,\AA\,.

Data reduction of the MOS spectra through to the point 
of producing rectified and wavelength calibrated 2D 
spectra was accomplished utilising the Gemini/GMOS 
packages in IRAF as described in \cite{Bridges06}. From 
the CuAr arcs, wavelength calibrations with residuals 
$\sim$0.1\,\AA\, were achieved. The extraction 
of the target GC spectra from each 2D spectrum was undertaken 
using the APALL task in the APEXTRACT package using an optimal 
(variance weighted extraction). Flux calibration was achieved
using GMOS longslit observations of the flux standard star
LTT3864 made during the same semester with identical
observational set-up.
 
The extraction of the NGC 3923 diffuse light spectra from 
the MOS slitlets was undertaken using a custom IDL script 
implementing the algorithm described below.


\section{Method}
Modern multi-object spectroscopic instruments such as GMOS 
are capable of taking spectra of around 40 objects 
simultaneously over fields of view of 
around 6$\times$6 arcmin$^{2}$. Each MOS slitlet is typically 
around 1 arcsec wide by a few (4-10) arcsec long. Generally 
the GCs are unresolved from the ground producing spectra 
smeared over $\sim$1 arcsec (in the spatial direction) for typical seeing 
values. The rest of the slitlet collects background (sky + galaxy)
photons used to background subtract the GC spectrum. 

\begin{figure} 
   \centering
   \begin{turn}{0}
   \includegraphics[scale=0.575]{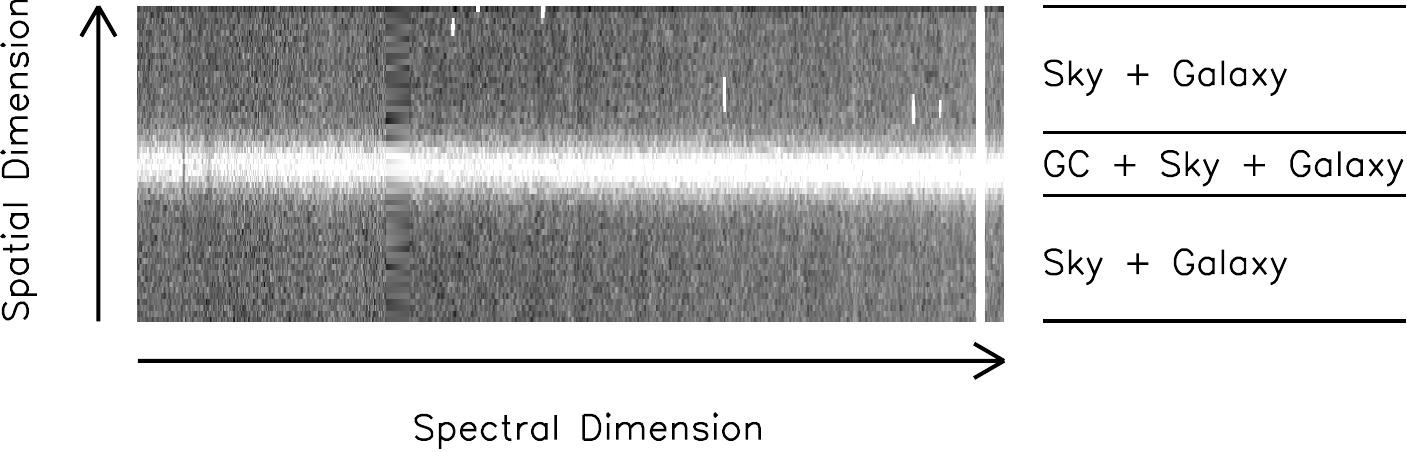}
   \end{turn}
    \caption{Example of a typical 2d MOS spectrum, showing
    target spectrum (in this case a GC spectrum) and sky regions
    typically used to sky subtract the target spectrum. For
    typical MOS GC studies these sky regions can contain
    useful amounts of host galaxy diffuse light. }
   \label{fig:example_2d}
\end{figure}

For the typical field of view of the current generation
of MOS instruments and the distance and 
size of the target galaxies being investigated by these studies, 
it can be seen that significant amounts of target galaxy flux 
are contained in the MOS slitlets being used to study the GCs 
(See Fig.~\ref{fig:mos_field}). The flux incident on each slitlet
can then be thought of as a sum of contributions from the 
target GC, the actual sky background (both atmospheric and 
extra-galactic) and a contribution from 
the diffuse light of the target galaxy. Fig.~\ref{fig:example_2d}
shows an example of a typical 2D MOS spectrum, showing the 
regions containing the different contributions to the flux. 
If the instrument has a sufficiently large field of view, it is 
possible to place a slitlet at large galactocentric radii 
where the spectrum measured would be essentially a 
pure sky spectrum, uncontaminated by galaxy flux. This sky spectrum 
could then be subtracted from the background regions of each 
of the GC slitlets to produce a sky-subtracted galaxy diffuse light 
spectrum for each of the 
individual slitlets. We have successfully utilised this approach 
in a study similar to this one involving the GC system of 
NGC 524, finding that slitlets located beyond 5R$_e$ provide 
an adequate sky spectrum.

In practice the FOV of the current generation of MOS 
instruments is often too small to allow a slitlet 
to be located at large enough galactocentric radii to 
produce a ``pure'' sky spectrum of the accuracy required 
(at least for stellar population studies).


Here we present an algorithm that allows the calculation of the
sky spectrum even in cases with small FOV, by combining the 
spectroscopic dataset with wider field imaging data.

\subsection{Algorithm}

\begin{figure} 
   \centering
   \begin{turn}{0}
   \includegraphics[scale=0.625]{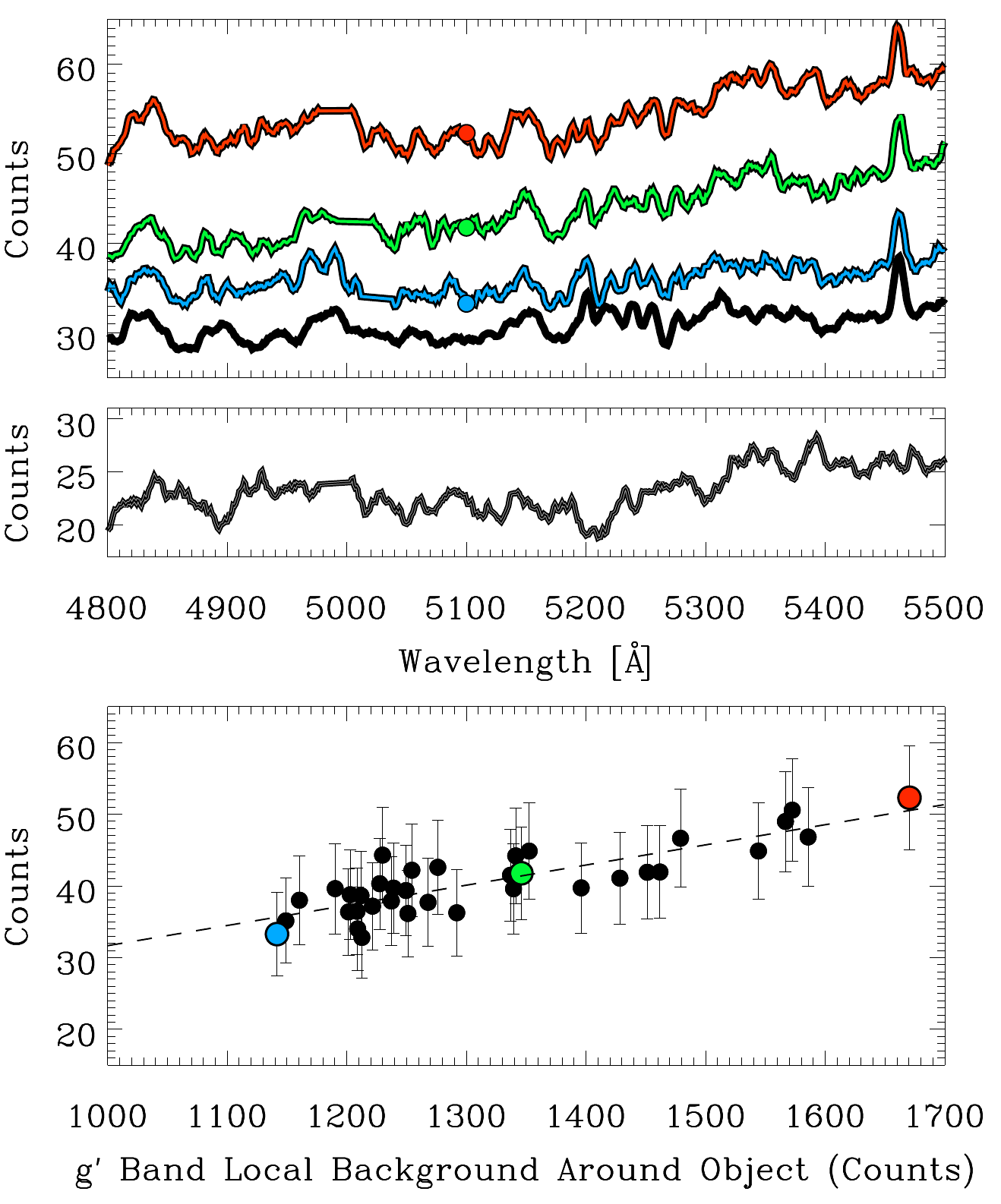}
   \end{turn}
    \caption{Upper Panel: Upper three spectra are the ``sky + galaxy'' spectra
    extracted from slits covering the range of galactocentric radii
    covered in this study. The fourth spectrum (black line)
    shows the extrapolated ``pure'' sky spectrum using the algorithm
    described in the text. 
    Mid Panel: This spectrum
    is the result of subtracting the extrapolated sky spectrum from the
    uppermost ``sky + galaxy'' spectrum. Redshifted (z=0.006) absorption lines
    due to H$\beta$, Mg$b$, Fe5270 and Fe5335 are clearly visible. 
    Lower Panel displays a
    typical result of the extrapolation process used to determine the 
    actual pure sky spectrum in this case for the pixel located at 
    5100\,\AA\,. The position of the 3 points from panel
    A is also displayed here for instructive purposes, as are the 
    measured number of counts and statistical error at this 
    pixel (5100\,\AA\,) for the other
    34 slitlets (black circles) in this mask.}
   \label{fig:galaxy_spec}
\end{figure}

\begin{figure} 
   \centering
   \begin{turn}{0}
   \includegraphics[scale=0.575]{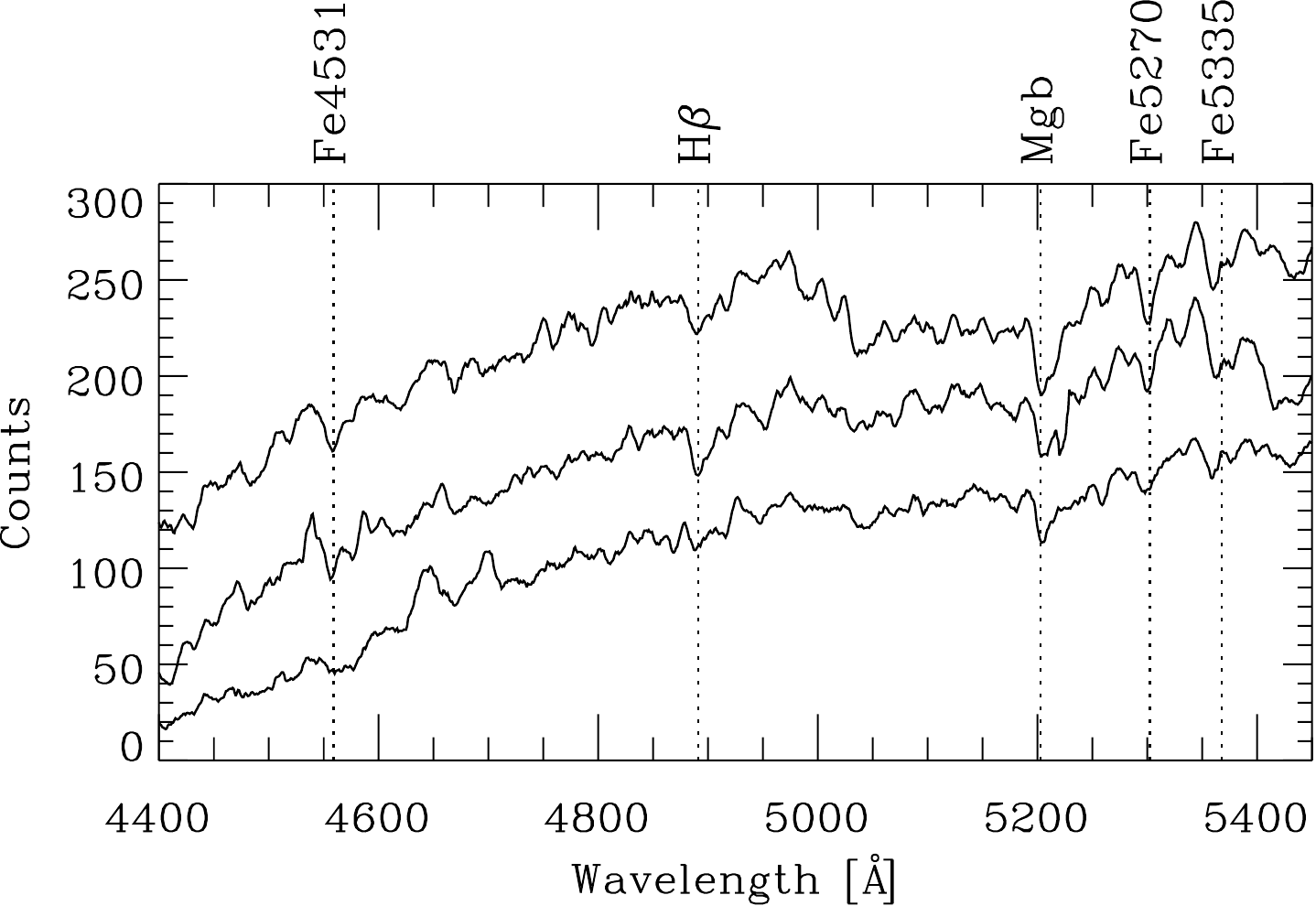}
   \end{turn}
    \caption{Representative diffuse light spectra for NGC 3923
    measured using our new technique. These spectra are the
    results of coadding 16 individual exposures,
    they have been smoothed to the Lick/IDS resolution. Redshifted
    absorption lines due to several species are present.}
   \label{fig:gal_spec}
\end{figure}

\begin{itemize}

\item  The spectra are reduced as normal to the point where
	  the individual spectra are cut out, rectified and wavelength
	  calibrated. See \cite{Bridges06} for more details of the
	  reduction used for the data presented here.
	  
\item  A sky + galaxy spectrum is produced for each slitlet. The
	 spectra are produced using the usual method for calculating
	 a sky spectrum from a 2D slitlet (i.e the spectrum used to
	 sky subtract a 2D frame). Fig.~\ref{fig:galaxy_spec}  
	 displays three such sky + galaxy spectra for a range of 
	 galactocentric radii; as is expected those spectra taken from 
	 slits which lie closer to the galaxy centre contain more flux.

\item Each spectrum is rebinned onto a uniform wavelength scale.

\item The wider field pre-imaging is combined to produce a master
	 wide field image of the target galaxy. In this case we make
	 use of our g$^\prime$ band imaging as this best matches the wavelength
	 coverage of our spectra, though we note that if our i$^\prime$ band 
	 imaging is used instead we obtain line indices which differ
	 by less than 0.1\,\AA. We caution however that care should be taken to
	 ensure that observed optical light profile being used should
	 match that of the spectral region being examined. i.e. that the
	 use of an IR image to set the scaling for optical spectra is
	 probably unwise because of the different photometric profiles displayed
	 by galaxies in the different wavelength regimes.

\item Using the master image of the field, the local background 
	 in an annulus around each of the target GCs is calculated.
	
\item The actual sky background number of counts in the image is measured
	 for a region located at large galactocentric radii. Experience 
	 with the data for NGC 3923 showed that this value was always
	 very similar to the local background measured for the GC with the largest 
	 galactocentric distance.

\item For each wavelength pixel of the rebinned spectra the number of counts 
	 is plotted against the local background around the 
	 corresponding object determined from the wide field image. 
	 (See Fig.~\ref{fig:galaxy_spec}  for the pixels corresponding to
	 5100\,\AA\,). As can be seen in the lower panel of 
	 Fig.~\ref{fig:galaxy_spec} the correlation of local background
	 with number of counts in the spectra is reasonably good. This
	 correlation can then be fit with a simple straight line and 
	 extrapolated to the value calculated for the sky background
	 at large radius from the wide field image. In the example in 
	 Fig.~\ref{fig:galaxy_spec} the value of the sky brightness
	 measured at large radii to represent the uncontaminated
	 sky value was around 1000 counts
	 in the g' band, leading to an estimate of the
         ``pure'' sky spectrum having around 32 counts in the pixel
	 located at 5100\,\AA\,.
	 
\item  When repeated for each wavelength pixel, this procedure 
	 results in a spectrum that is a good 
	 approximation of a ``pure'' sky spectrum. Note that the 
	 produced sky spectrum is only accurate in the regions where a 
	 significant number of the spectra overlap 
	 in wavelength coverage. In the case of our GMOS data this 
	 restricts our analysis of the galaxy spectra to the 
	 wavelength range ~4400 - 5500\,\AA\, where the majority 
	 of the 37 input spectra have coverage.

\item Each of the 37 rebinned sky + galaxy spectra are then sky 
         subtracted using the ``pure'' sky spectrum, leaving 37 
	 spectra that are essentially pure galaxy spectra. 
	 The spectra measured from individual exposures
	 can then be combined in the usual manner; examples of
	 3 of the spectra produced by combining 16 individual
	 exposures can be seen in Fig.~\ref{fig:gal_spec}.\\

In cases where wider field imaging is not available it is possible
to obtain similar results using an iterative procedure.
In this case an initial guess for  the sky brightness 
assumed to represent pure night sky is made (by 
extrapolating the observed galaxy profile for instance) and the entire
procedure from steps 6 to 8 can then be iterated with the assumed
night sky brightness being varied until sky line residuals in the final 
diffuse light spectra are minimised.

\end{itemize}


\subsection{Kinematics}

GC radial velocities were measured using the FXCOR
task in the RV package of IRAF. We did not observe radial
velocity standard stars as part of this investigation due
to the extra overhead in observing time this
requires. For templates we therefore use spectra
from the simple stellar population library of 
\citet{Vazdekis99}. These templates span a 
range of age of 1-18 Gyr and a metallicity range from 
[Fe/H] = +0.2 to -0.7. To this sample we add 6 
stellar spectra from the \citet{Jones97} library to extend
the metallicitiy range covered by the templates to
[Fe/H] $<$ -1.5. This wide range of age and 
metallicity helps to minimise the effects of template 
mismatch on the measured velocities. We use a 3 sigma clipped
mean of the velocities derived from these templates as the
final velocity for each GC. The errors are estimated from
the mean of the errors measured by FXCOR for those 
velocities not removed by the clipping procedure.

Objects with velocities in the range 1200-2400kms$^{-1}$ 
are assumed to be associated with NGC 3923 and
were classified as GC. We find that 29 out of 37 (78$\%$) 
objects observed with our MOS mask have velocities within this
range. Of the remaining objects, 2 were z$\sim$0.3 
galaxies, one was a z$\sim$1.3 quasar and 5 were Milky Way 
stars.

The recessional velocities of the NGC 3923 diffuse light 
spectra were measured using the publicly available IDL
implementation of the pPXF method \citep{pPXF}. This method 
was chosen as it allows a simultaneous measurement of the 
radial velocity and velocity dispersion from the spectrum, both of 
which are important in studies of the diffuse light of 
galaxies. In this instance the spectra were fit in pixel 
space over the wavelength range 4800-5400\,\AA\, using 
the same templates as the GCs. The final velocity and 
velocity dispersion for each slitlet is taken to be the 
mean of the velocities/velocity dispersions determined 
from 100 Monte-Carlo resimulations of the input spectra 
with added photon noise. The error in the measured velocity 
is taken to be the dispersion in the 100 best fit velocities. 
As a check of consistency we remeasured the recessional 
velocity of the diffuse light spectra using the FXCOR task 
in IRAF and found no significant differences between the 
two methods.

\subsection{Line Indices}
By binning all of our 37 diffuse light spectra together
we have produced one spectrum of sufficient signal-to-noise
(S/N $>$ 60 per \AA) to measure accurate absorption line indices.
Since NGC 3923 shows evidence of slight rotation 
(Fig.~\ref{fig:gal_rot}) it is necessary to remove the effects of 
rotation on the spectra by correcting the spectra to some common 
velocity before coadding them. In this instance the velocity of 
each of the 37 spectra was measured as described previously, 
then using the DOPCOR task in IRAF each of the spectra was 
de-redshifted to zero redshift before combining.

For ease of comparison with other studies we have chosen
to use the Lick/IDS system when examining the stellar
population present in this master spectrum and in our GC
sample. There are several steps that must be taken to ensure 
that our measured indices are securely on the Lick/IDS 
system \citep{Worthey97,Trager98}.
The first is to convolve our spectra with a 
wavelength dependent Gaussian kernel to reproduce the
variable Lick/IDS resolution of 9-11\,\AA\,.
We then measure the Lick/IDS indices for our fluxed spectra
using the wavelength definitions from 
\cite{Worthey97}, \cite{Trager98} and the INDEXF index measuring
code of N Cardiel\footnotemark 
\footnotetext{Available at: \\
http://www.ucm.es/info/Astrof/software/indexf/indexf.html}. 
Uncertainties on the measured 
indices are a combination of errors introduced by photon
(both target and sky) noise and errors in the measured redshift 
of the spectra (essentially negligible in all cases).

The line indices for the spheroid spectra must also
be corrected for the smearing effects of the line-of-sight
velocity distribution (LOSVD). This effect is 
entirely negligible for GC spectra with velocity dispersions 
of the order a few kms$^{-1}$, but cannot be ignored for
galaxies with velocity dispersions over $\sim$100 kms$^{-1}$.
As our measured diffuse light spectrum has a velocity dispersion
of around 200 kms$^{-1}$ this step is important.
We correct for this effect using the procedure of 
\cite{Kuntschner04} which can also
be used to correct line indices for the additional
influences of the higher order LOSVD terms h${_3}$ and
h${_4}$. \cite{Kuntschner04} provides a new parameterisation
for the LOSVD line strength corrections derived by 
determining the difference in measured Lick/IDS line 
strengths measured from template spectra before 
and after convolving with various choices of $\sigma$, 
h${_3}$ and h${_4}$.
For NGC 3923 the correction for h${_3}$
and h${_4}$ is not implemented due to the effect of the
correction given the measured values of $\sigma$, h${_3}$ and
h${_4}$ being minimal relative to the other errors.

The final step required to firmly fix our measurements
to the Lick/IDS system is to measure standard stars
from the Lick library using our observational
set-up and compute offsets between the two. Although 
we did not observe Lick standard stars directly during this project, 
due to the prohibitive cost in observing time, data
collected as part of GS-2003B-Q-63 to calibrate 
GMOS-S onto the Lick system were kindly made 
available by Bryan Miller. We have used a set
of 28 stars from this sample, taken with an almost
identical set-up to our MOS observations and used
them to derive offsets for GMOS-S to the Lick
system. We have applied these offsets to our measured
line indices presented in Table \ref{tabobs}. We note that while useful for comparison
between observations taken using different instruments,
or for comparing measured indices to models based 
on the Lick/IDS system e.g. \citet{Thomas03,Thomas04}, this step
is not essential for a simple first order differential
comparison between two sets of observations made with the
same instrument, such as that described here.

Using the procedure outlined (neglecting the higher order LOSVD
corrections) we have measured 
Lick/IDS line strength indices for all 29 of our confirmed
GCs, as well as our single coadded spheroid diffuse light spectrum.
This single diffuse light spectrum is located at a luminosity weighted
distance of 3 R$_e$ (throughout we use R$_e$=43.8 arcsec 
measured in the J-band \citep{Jarrett03}, though we note that V, R 
and I band estimates from \citet{Bender88} are essentially identical). The 
S/N of our GC spectra range from $\sim$ 11-56 per \AA\,
measured in the region 5100-5150 \AA\,, giving errors in the
H$\beta$ index of 0.12-0.72 \AA\,.

\section{Results}

Table~\ref{tabobs} presents velocities and Lick/IDS line strength 
indices for all 29 of our confirmed GCs, as well as the single coadded
diffuse light spectrum, and a series of GC composite indices binned by 
g$^\prime$-i$^\prime$ colour. In addition to examining the GC spectra we have 
extracted 37 diffuse light spectra from our MOS exposure (see 
Fig.~\ref{fig:gal_spec} and Table~\ref{tabkin}) with signal-to-noise 
ranging from 5 to 12, which is sufficient to measure velocities in all 
cases and velocity dispersions in most cases. 

\subsection{Kinematics}

A more detailed discussion of the kinematics of NGC 3923 and 
its GC system will be presented in Norris et al. (in prep), where 
we include GC velocities from additional masks at larger radii 
obtained using the Nod and Shuffle technique \citep{vaetvient}. 

In this section we limit ourselves to examining the spheroid 
of NGC~3923 for rotation and in Section 4.2 to comparing 
the velocities measured for the spheroid of NGC~3923 and 
those measured for the GCs examined in the same slits.

\begin{figure} 
   \centering
   \begin{turn}{0}
   \includegraphics[scale=0.9]{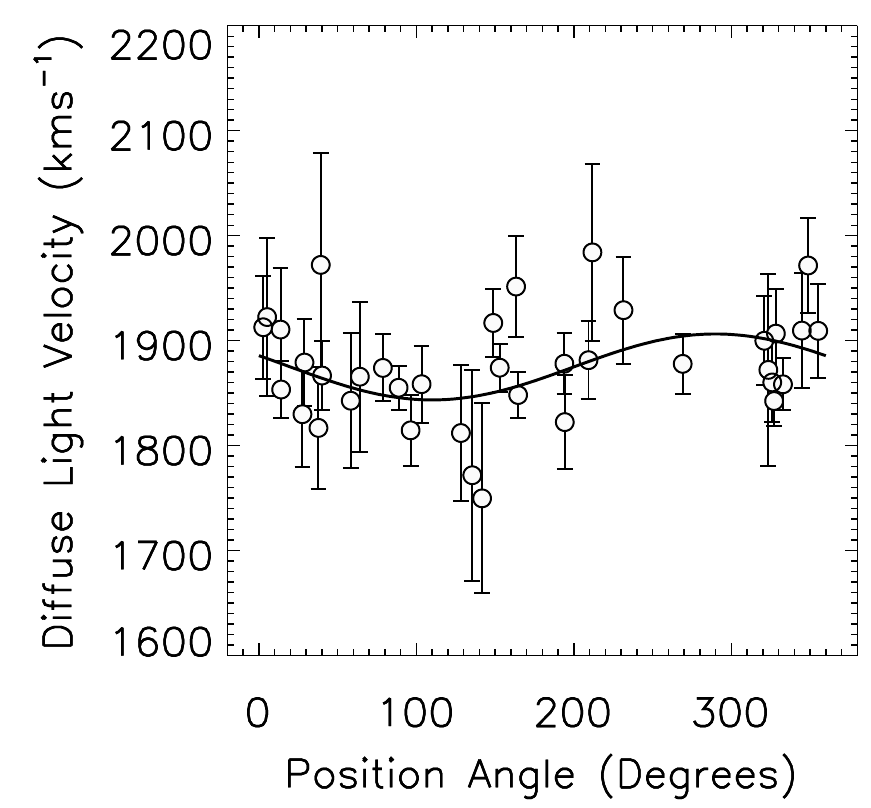}
   \end{turn}
    \caption{Velocity vs. Azimuthal angle for
    NGC 3923 spectra extracted from MOS slitlets (circles).
    The solid line is the best fit flat rotation curve indicating
    rotation amplitude of 31 kms$^{-1}$ $\pm$ 13 with
    position angle 290$^{\circ}$ $\pm$ 19 (north through east).
    }
   \label{fig:gal_rot}
\end{figure}

Here we present the results of our search for evidence of 
rotation in the diffuse stellar halo of NGC 3923 
(see Fig.~\ref{fig:gal_rot}) by carrying out a non-linear least 
squares fit to the equation:
\\
\\
V($\theta$) = V$_{rot}$sin($\theta$ - $\theta$$_{0}$) + V$_{0}$ 
\\
\\
where V(${\theta}$) is the velocity measured for the 
diffuse light in each slitlet, V$_{rot}$ is the 
amplitude of the projected rotation velocity, V$_{0}$
the systematic velocity of NGC 3923, $\theta$ the azimuthal
angle of each slitlet relative to the galaxy major axis
and $\theta$$_{0}$ is the position angle of the line of 
nodes. This approach determines the best-fitting
simple flat rotation curve (see \cite{Zepf2000} for details).

We find evidence for rotation along the major axis of
NGC 3923, with the best fitting amplitude being 31 $\pm$ 
13 kms$^{-1}$ along a position angle of 290 $\pm$ 19$^\circ$ 
where the major axis is approximately at PA = 315$^\circ$. 
This result is in good agreement with that of  \cite{Koprolin2000}
who found a small (53 $\pm$ 13 kms$^{-1}$) but non 
negligible major axis rotation for NGC 3923. It is however
different to the results of  \cite{Carter98} who found 
that the inner 25 arcsec of NGC 3923 showed no rotation 
on its major axis but had minor axis rotation of amplitude 
$\sim$ 20 kms$^{-1}$. The difference in conclusions between the
different studies may be attributed to the different radial ranges over
which they are sensitive; our method provides a measurement
of the rotation of the galaxy at a radius at least a factor of two
greater than the largest radial point available to either 
\cite{Carter98} or \cite{Koprolin2000}.

These measurements of the kinematics of NGC 3923 cover 
a range in effective radii of between 2 and 4 R$_{e}$, 
demonstrating the great power which this approach has for 
illuminating the kinematics of the outer regions of galaxies.

\subsection{Reliability of Velocities}
\begin{figure} 
   \centering
   \begin{turn}{0}
   \includegraphics[scale=0.9]{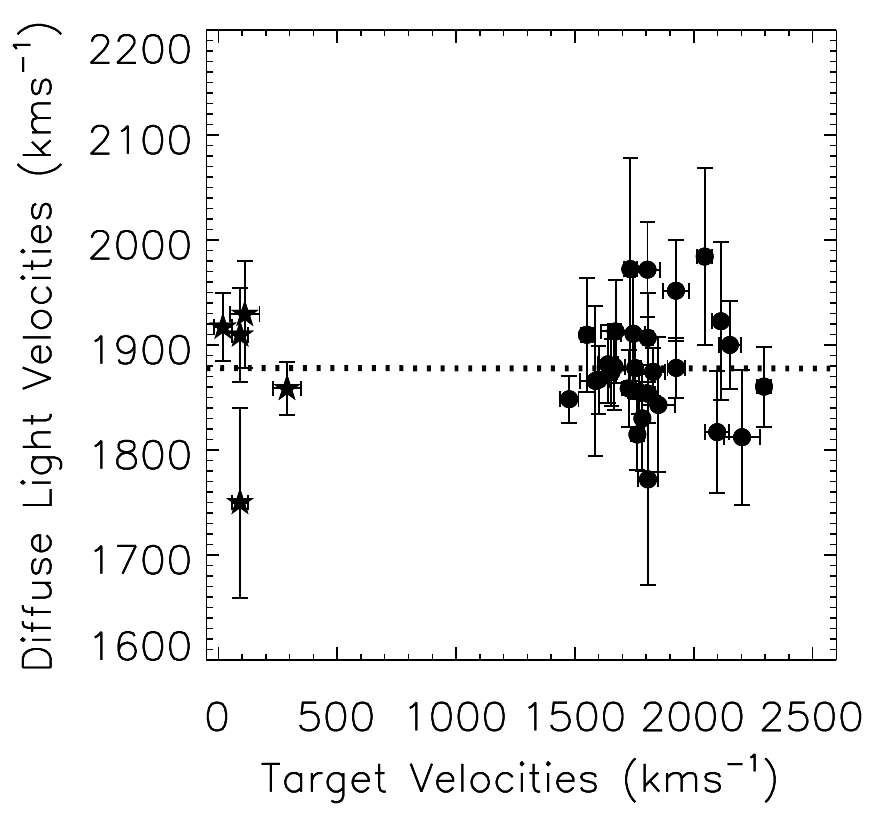}
   \end{turn}
    \caption{Measured velocities
    from NGC 3923 diffuse light spectra vs. the measured velocities
    of the targets observed in the same slits. Stars indicate
    targets classified as Milky Way stars, filled circles those targets 
    classified as GCs belonging to the NGC 3923 system. Two z$\sim$0.3 
    galaxies and a z$\sim$1.3 quasar are excluded for clarity. The dotted 
    line shows a linear fit to the data, indicating no correlation
    between the velocities measured for the targets and the galaxy diffuse
    light.}
   \label{fig:gal_kin}
\end{figure}

One concern with our procedure for extracting diffuse host
galaxy spheroid light from the same slitlets as the target GC
spectra is that the spheroid spectra could be contaminated
by flux from the target GCs. In general this should not be
of major concern due to the careful selection of the sky
regions of the 2D spectrum. As a test of this assumption
in Fig.~\ref{fig:gal_kin}, we plot the measured diffuse light velocity
against the target velocity (GC, Milky Way star, background 
galaxy or QSO) from the same slitlet. In Fig.~\ref{fig:gal_kin} the filled circles
are target objects classified by their velocities as GCs, filled stars
are objects classified as Milky Way stars, the three background
objects (two z$\sim$0.3 and a z$\sim$1.3 QSO) are omitted
from the plot for clarity, although they are included in the fit
which produces the dashed best fit trend line (which has slope 
equal to zero within the errors).
As can be seen clearly, there is no
correlation between the velocities measured for the target
spectra and those measured for the diffuse light; in particular
even for the brightest target objects (generally the MW stars)
the measured velocity for the spheroid of NGC 3923 is still
consistent with the recessional velocity of NGC 3923.
This indicates that contamination of the diffuse light 
spectra by the target object spectra is negligible.


\subsection{Stellar Populations}
In this section we examine the stellar populations of the GCs and the
diffuse light of NGC 3923. To begin we shall discuss some of the 
qualitative properties of the stellar populations through examination 
of age-metallicity diagnostic plots. We shall then
examine the data more quantitatively by presenting the 
ages, metallicities and $\alpha$-element abundances derived through 
$\chi^2$ fitting of the data to stellar population models. 

\subsubsection{Index-Index Plots}

\begin{figure*} 
   \centering
   \includegraphics[scale=1.0]{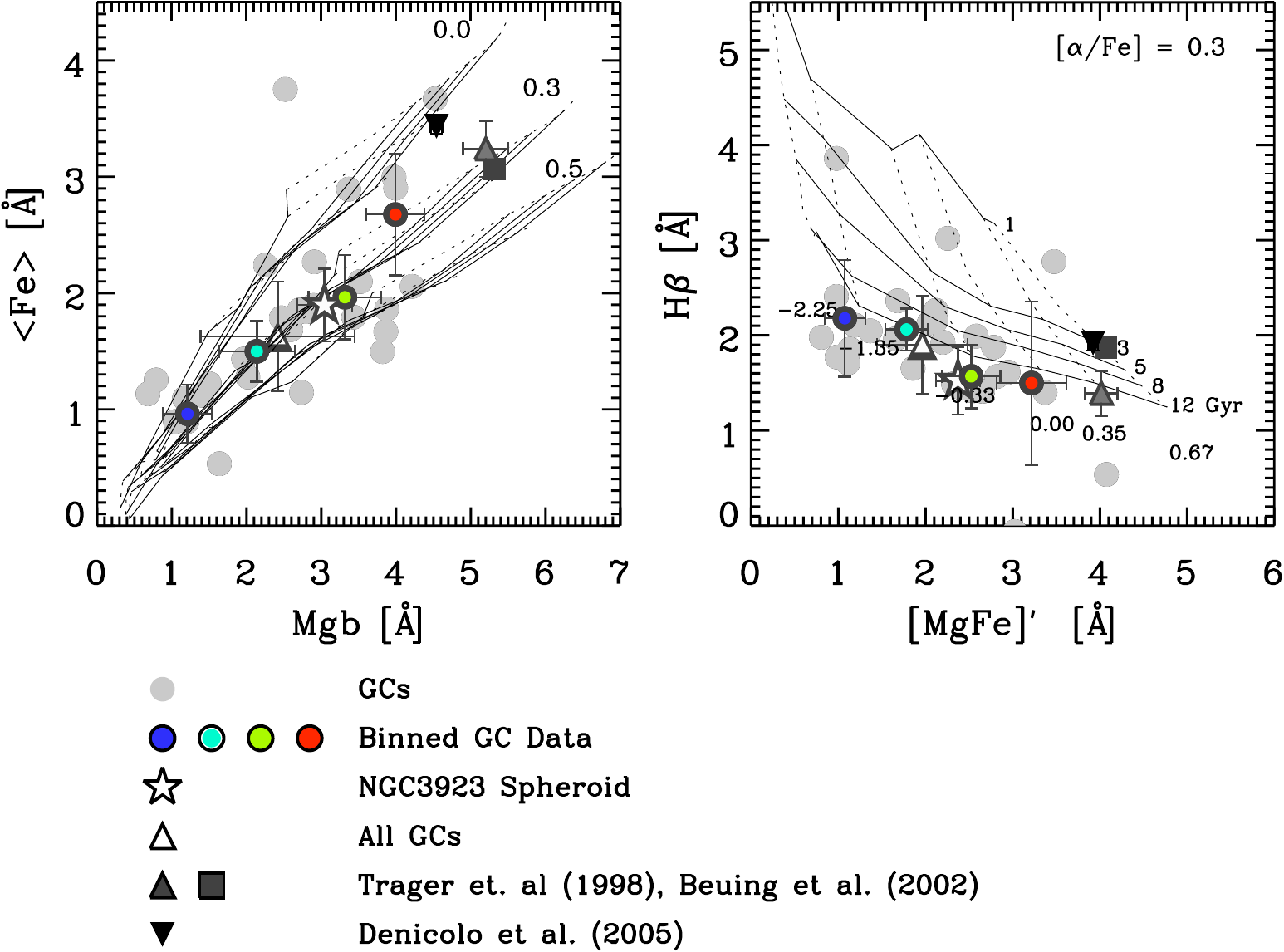}
    \caption{Left Panel. Comparison of the [$\alpha$/Fe] ratios of
    the diffuse light and GC population of NGC 3923 through the use
    of an Mg$b$ vs. $<$Fe$>$ diagram. Light grey circles are the
    29 GCs measured here (error bars omitted for clarity).
    The coloured circles are the result of binning the 29 GC data points
    into 4 bins of 7/8 GCs per bin by g$'$-i$'$ colour, the error bars show the 
     1$\sigma$ scatter in each bin. The white triangle is the 
    result of binning all 29 GC spectra into one bin, again the error bar 
    represents the 1$\sigma$ scatter in the bin. The grey triangle, square 
    and black triangle are the central NGC 3923 values from 
    \citet{Trager98}, \citet{Beuing02} and \citet{Denicolo05} respectively. The white star shows
    the values measured from the single high S/N coadded spectrum of the
    diffuse light of NGC 3923. Overplotted are models by \citet{Thomas03,Thomas04}
    with abundance ratios [$\alpha$/Fe] = 0.0, 0.3, 0.5, the models
    have ages 3-12 Gyr and metallicity [Z/H] = -2.25, -1.35, -0.33, 0.0,
    +0.35 and +0.67. Right Panel. Age-Metallicity diagnostic plot. Symbols as in 
    left panel. Models are overplotted for a fixed value of  [$\alpha$/Fe] = 0.3
   }
   \label{fig:indices}
\end{figure*}

In the left panel of Fig.~\ref{fig:indices} we present an Mg$b$ vs. $<$Fe$>$ 
plot\footnotemark  
\footnotetext{$<$Fe$>$ = (Fe5270 + Fe5335)/2. \citep{Gonzalez93}}
for the 29 GCs as well as the NGC 3923 diffuse light spectrum 
measured in this study. Model predictions from \citet{Thomas03,Thomas04}
are overplotted for [$\alpha$/Fe] = 0.0, 0.3 and 0.5, ages 1-12 Gyr and 
metallicity running from -2.25 (bottom left) to +0.67 (top right). This choice of indices
produces a figure which is insensitive to age but can be used to constrain
the metallicity and $\alpha$-element abundance ratios \citep{Kuntschner02}. 
As can be seen the GCs (light grey circles)
have metallicities ranging from -1.7 to +0.5 and appear to
have a wide range of $\alpha$-element abundance ratios. 
When the GC data is binned
by colour (coloured circles) the behaviour of the binned data 
becomes more systematic, with the points showing a clear preference 
for a constant $\alpha$ element abundance ratio of $\sim$0.3, even as the 
metallicity of the binned data decreases from around 0.0 to -1.5.
These bins cover the range of colour displayed by the majority 
of extragalactic GCs, and have mean g$'$ - i$'$ colour 
of 0.82, 0.99, 1.09 and 1.19 (See Table \ref{tabobs}). The position of the mean colour
of each bin relative to the peaks of the blue and red GC distributions
can be seen in Fig.~\ref{fig:phot}. In particular the mean colour
of the third bin can be seen to correspond closely to that of the peak
of the red GC distribution. If all the GC data 
are combined to produce a composite GC data point (white triangle);
the combined GC spectrum has a metallicity of $\sim$ -0.8 and an 
abundance ratio consistent with 0.3 within the (significant) errors. It is
interesting that both the central regions of NGC 3923 measured by
 \citet{Trager98} and \citet{Beuing02} (grey triangle and square) and the 
NGC 3923 diffuse light spectrum measured in this study (white star) also
lie along the same [$\alpha$/Fe] = 0.3 track. However the fact that the
data point from \citet{Denicolo05} (black triangle) for the central
R$_e$/8 of NGC~3923 is offset relative to the other two observations 
illustrates the magnitude of possible systematic errors in this type of
study.   The difference in metallicity
between the inner nuclear data and our large radii diffuse stellar data 
point can be understood in terms of a radial metallicity gradient in the galaxy.
A further point of note is the close agreement between the mean of the GC 
spectrum and the NGC 3923 diffuse light spectrum which was measured
at the same projected radii.


The right panel of Fig.~\ref{fig:indices} presents an 
[MgFe]$'$ vs H$\beta$ age-metallicity plot\footnotemark 
\footnotetext{[MgFe]$'$  = $\sqrt{Mgb \times (0.72 \times Fe5270 + 0.28 \times Fe5335}$ \citep{Thomas03}} 
  with symbols as previously defined.
 Here it is necessary to choose a fixed value of [$\alpha$/Fe] for
 the grids and we have chosen 0.3 as this value is 
 close to that found for each of the populations described in the
 previous section. It can be seen that three GCs 
 appear to lie along grid lines consistent with young or intermediate
 ages, these objects are however generally among the lowest
 S/N objects in our sample. When other age-metallicity diagnostic
 plots (H$\gamma$ and H$\delta$ vs. [MgFe]') are considered as 
 well as the fact that similar numbers of objects lie below the grid 
 it appears that these objects may just be statistical outliers in a
 distribution centred on uniformly older ages.
 We shall return to this point when examining the results
 of the $\chi^2$ fitting approach.
 The binned GC data again
 behaves in a systematic manner, in which all the data points lie
 along a constant age line of around 12 Gyr, with metallicities
 consistent with those seen in the previous plot.
 The mean of the GC system as a whole and the NGC 3923 diffuse
 light point can also be seen to be very consistent in terms of age 
 and metallicity.

The close agreement in apparent age, metallicity and [$\alpha$/Fe]
between the NGC 3923 diffuse light spectrum measured at 3 R$_e$
and the GC system (especially the red GCs) at the same projected 
radii is tantalising. Such close agreement hints at a deep connection 
between the two populations exactly as would be expected if the
two populations formed coevally from the same material. 
This interpretation is however dependent on the assumption
that the GCs examined here are representative of the full GC population. 
We believe that our sample is 
a fair match to that present in NGC 3923 over the range of projected
radii studied, though only observation of a significantly
larger sample of GCs carefully chosen to match the full sample
in terms of colour and luminosity would allow us to determine this for
certain. \\

\subsubsection{Derived Stellar Population Parameters}

We derive ages, metallicities and $\alpha$-element abundances for
the stellar populations being studied here using the multi-index fitting
method of \cite{Proctor_Sansom} and \cite{Proctor04}. This technique compares the
measured Lick indices with SSP models, in this case the SSP models of 
\citet{Thomas03,Thomas04} (TMK04) which to date remain the only SSP models
in which the effects of $\alpha$-element abundance ratios on Balmer
line indices at low metallicity are included. For a more thorough description
of the different treatments of $\alpha$-element abundance ratios 
in modern SSP models see \citet{Mendel07}.

The procedure used to determine the best fitting SSP model is as
follows. The total set of measured Lick/IDS indices is compared to the 
TMK04 SSPs and a minimum $\chi^2$ fit obtained; simultaneously
a set of $\chi^2$ minimization fits are determined with each of the
indices omitted. The lowest total $\chi^2$ fit from this set is then chosen,
the necessary index removed and the process repeated until a 
stable fit is arrived at where no highly aberrant (3$\sigma$) indices remain.
In this instance all GCs displayed some indices which were 
sufficiently aberrant to be excluded from the fit, though all fits
utilised a minimum of 9 indices, most GCs are fit with at least 13
indices. Errors in derived parameters are determined via 
monte-carlo re-simulations of the input index measurements with 
their measured errors. The results of this procedure are displayed
in Table~\ref{tabres}.

In common with previous studies we find that the molecular
bands Mg$_1$ and Mg$_2$ appear depressed, an observation 
generally explained as being due to problems with flux calibration.
We have tested this explanation by examining the effect of using 
the flux calibration curves found from repeat observations of the 
same standard star (in this case LTT4364) observed for a different 
program using the same observational set up used here. We found that 
even when using curves derived from the same standard star the measured 
Mg$_1$ and Mg$_2$ indices of our GCs after flux calibration could vary 
systematically by as much as 0.01 mag, which is of the same order as
the deviations observed in our fitting procedure. Other indices were 
much less affected by the change in flux calibration, presumably due to 
the much narrower wavelength coverage of these indices. Accordingly 
Mg$_1$ and Mg$_2$, as well as Ca4227 which was similarly depressed, were
therefore removed from our fitting procedure. The CN indices tended to 
be enhanced relative to the models, in agreement with previous findings 
\citep{Beasley04_GC,Pierce3379} for GCs.
The Ca4668 index also tended to be unreliable and was therefore
excluded from the fits.

Our quantitative $\chi^2$ fitting procedure produces results in good
agreement with the more qualitative discussion in the previous
section. We find the GCs examined to have ages consistent at the 
one sigma level with old ages ($\ga$10 Gyr) in all but one case. The GCs 
exhibit a spread of metallicities from [Z/H] =  -1.8 to +0.35 and $\alpha$-element 
enhancement ratios generally consistent with $\sim$0.3 dex, although several
objects display [$\alpha$/Fe] of below zero or greater than 0.5, at the limits
of the models. From our sample of 29 GCs only one object (ID: 197) displays an age 
which is inconsistent with a mean age greater than 10 Gyr, however this
result is not significant being only a 1.5$\sigma$ deviation.
A more thorough analysis of the distribution of stellar population
parameters is somewhat limited by the small sample size, however to first
order we can state that the distribution of ages is consistent with a
single value of around 12Gyr plus measurement errors. The distribution
of metallicity however is inconsistent at the greater than 99.9$\%$ 
significance with having a single value plus measurement errors. We find 
that the distribution of metallicity measured is consistent 
with a mean of -0.65 dex with an intrinsic scatter of 0.5 dex plus 
measurement errors, assuming a gaussian distribution of metallicities. 
Similarly the distribution of [$\alpha$/Fe]
is also inconsistent with a single value plus measurement errors at the 
greater than 99.8$\%$ significance level, in this case we estimate an 
intrinsic scatter of 0.12 dex plus measurement errrors around a mean of 0.31 dex.

\begin{figure} 
   \centering
   \includegraphics[scale=0.65]{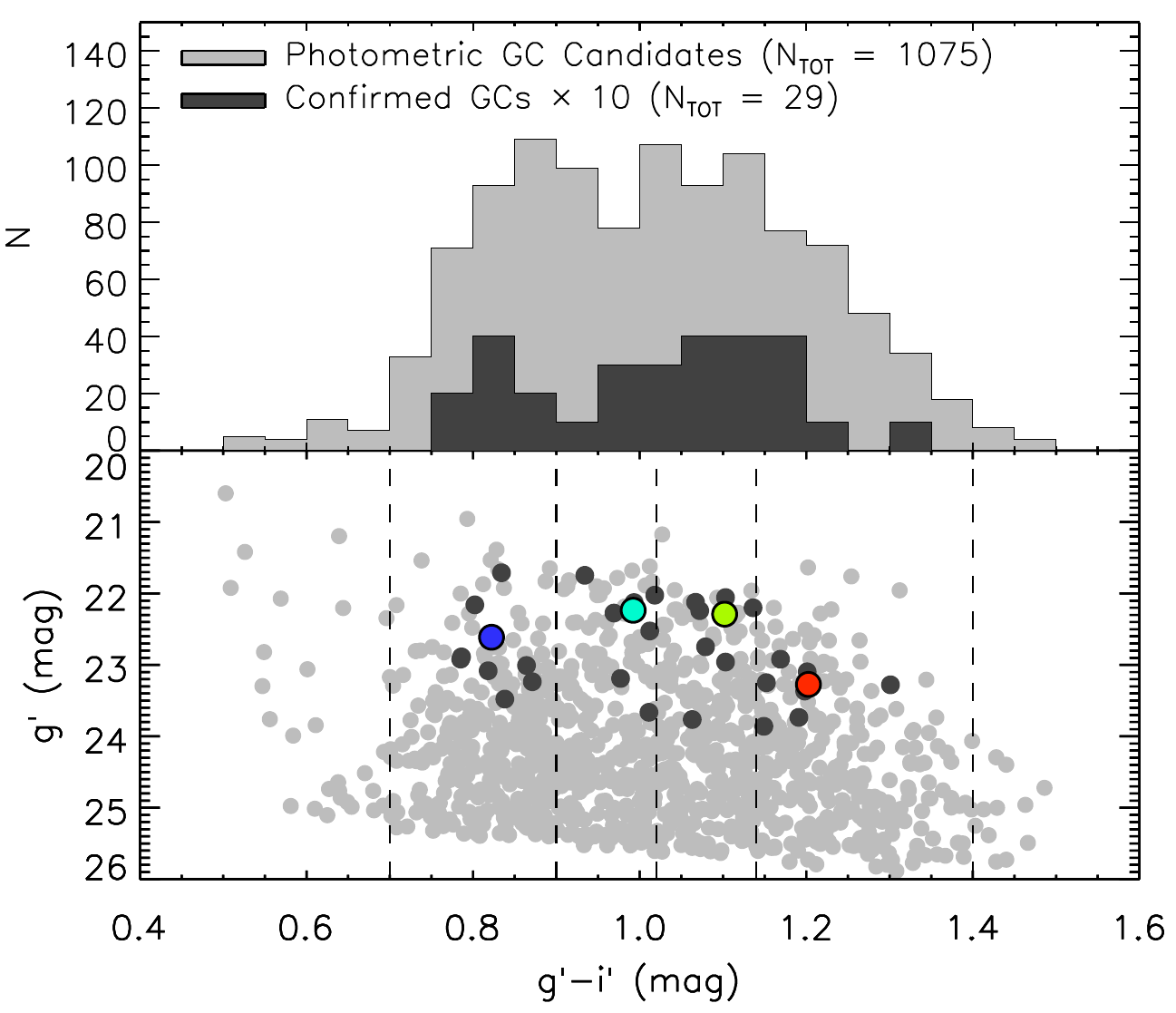}
    \caption{Top Panel. Light grey histogram shows the distribution
    of g$'$-i$'$ colours for photometrically selected candidate 
    GCs around NGC 3923 from our
    GMOS pre-imaging. Dark histogram (multiplied by a factor of 
    10 for clarity) shows the distribution
    of g$'$-i$'$ colours for our 29 spectroscopically confirmed GCs.
    Bottom Panel. g$'$-i$'$ vs g$'$ colour-magnitude diagram for our
    photometric (light grey circles) and spectroscopically 
    (dark grey circles) confirmed GCs. The coloured circles indicate 
    the error weighted mean values for each bin, the dashed lines
    indicate the arbitrary choice of bin ranges.}
   \label{fig:phot}
\end{figure}

We have repeated this analysis for our binned GC data and for the
line indices measured for the diffuse light halo of NGC 3923. We find that
the binned GC data again behaves in a manner consistent with the
more qualitative examination of the index-index plots. All bins 
have ages consistent with being older than 10 Gyr,
all four bins display a smoothly increasing metallicity with colour, ranging
from [Z/H] = -1.375 to -0.125, and all display [$\alpha$/Fe]  consistent with
0.3. The considerably larger errors present in the measured parameters for 
the reddest colour bin, are a result of this bin being 
composed almost entirely of lower luminosity and hence lower S/N objects. 
Apparently through chance selection this bin contains GCs of considerably 
lower lumnosity than the average for GCs examined spectroscopically 
here (see Fig.~\ref{fig:phot}). This results 
in a mean  g$'$ magnitude fully half a magnitude below any of the other bins;
together with the effect of error weighting the indices when combining 
them this explains why this bin is significantly lower S/N than the others.

Our ``Super GC" bin, containing the error weighted indices of all 29 GCs
unsurprisingly predicts ages, metallicities and $\alpha$-element abundances
in good agreement with those of the colour binned GCs. We derive an old
age for the GC population as a whole, a metallicity [Z/H] = -0.8 and an
$\alpha$-element abundance ratio which is consistent with 0.3.

Turning our attention to the measured parameters for the
diffuse light of NGC 3923 at $\sim$3R$_e$ we again find old ages,
metallicity [Z/H] of -0.325 and [$\alpha$/Fe] of 0.34. These results are
entirely consistent with that found for the red bin of GCs
(see Fig.~\ref{fig:stellarpops}), those
GCs displaying g$'$-i$'$ colours in the range 1.02 to 1.14.


\subsubsection{Stellar Population Parameter Corellations}

\begin{figure*} 
   \centering
   \includegraphics[scale=1.0]{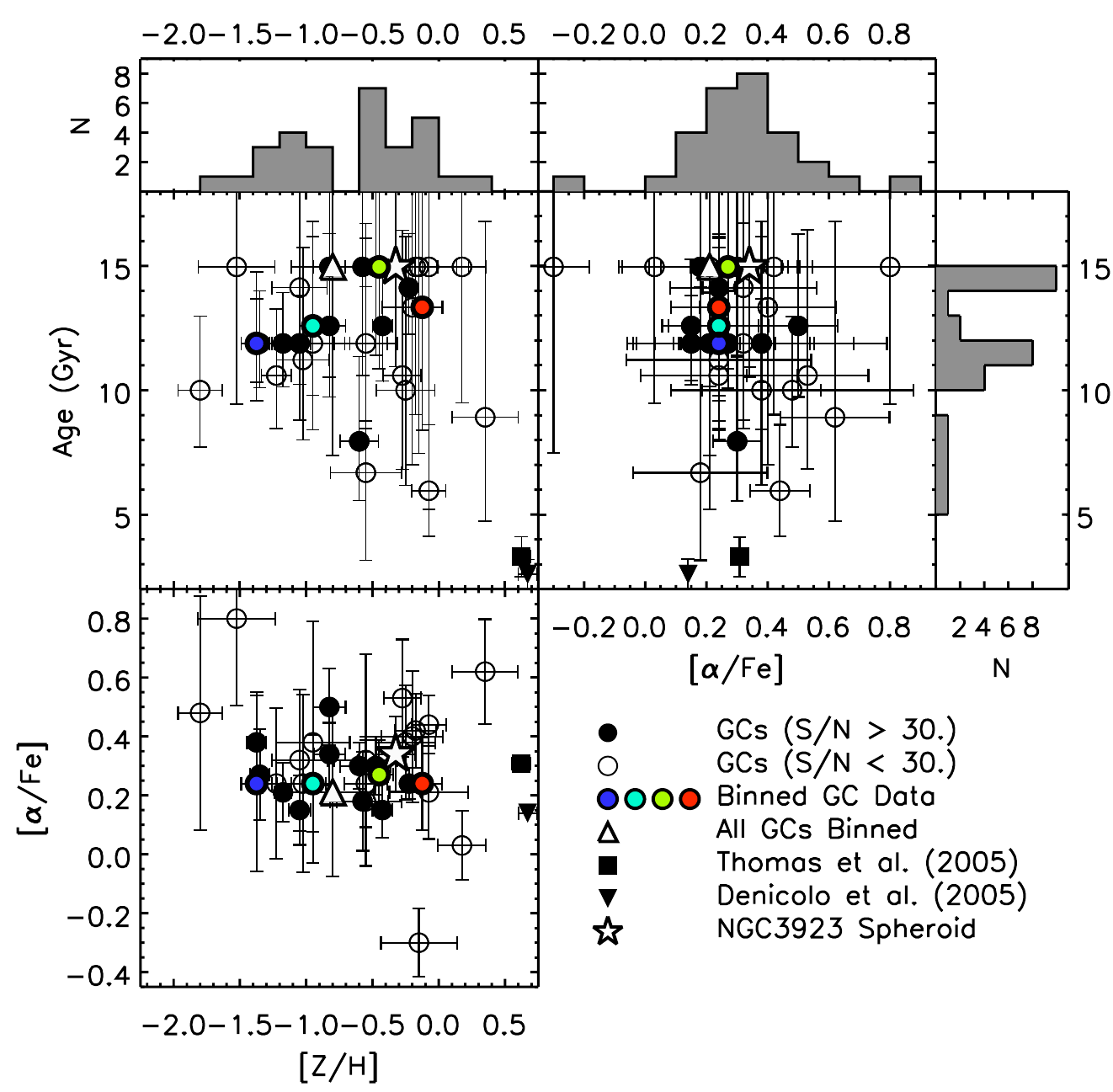}
    \caption{Ages, metallicities and [$\alpha$/Fe] ratios measured for
    the NGC 3923 GCs (filled circles have S/N per \AA $>$30, unfilled
    have S/N $<$ 30 per \AA), GC data binned by colour (coloured circles),
    NGC 3923 central regions from \citet{Thomas05} (filled square), 
    \citet{Denicolo05} (filled triangle) and  
    NGC 3923 spheroid (unfilled star).  }
   \label{fig:stellarpops}
\end{figure*}

In Fig.~\ref{fig:stellarpops} we search for correlations between the
ages, metallicities and alpha-element enhancements measured
for the various stellar populations using the $\chi^2$ fitting
procedure outlined previously.

One interesting observation is that we see no evidence for 
the observed trend of decreasing [$\alpha$/Fe] with increasing
metallicity seen in the GC systems of some galaxies 
(e.g. NGC 3379 \citep{Pierce3379}, NGC 4649 \citep{Pierce4649}), 
but not in others (NGC 1407 
\citep{Cenarro07}, VCC1087 \citep{Beasley06}).

Another interesting feature visible in our higher S/N data (filled circles) 
is that the \cite{Thomas03,Thomas04} models tend to introduce
a slight trend of increasing age with increasing metallicity. This
trend, which has been previously observed in the NGC 1407
system \citep{Cenarro07} and the MW GC system 
\citep{Cenarro07,Mendel07} appears to be a peculiarity of the
\cite{Thomas03,Thomas04} models; \cite{Mendel07} found
that the Lee $\&$ Worthey \citep{LeeWorthey05} 
and Vazdekis \citep{Vazdekis07} SSP models do not display
this trend. However as the age determinations at the lowest metallicities are 
already highly uncertain due to the possible effects of horizontal 
branch morphology, this effect is not significant and does
not affect our conclusions.


\section{Discussion}

In summary, we have obtained spectra of 29 GCs associated with 
NGC 3923, as well as 37 spectra of the diffuse light at radii 
of 2-4 R$_{e}$. By combining these 37 diffuse light spectra we 
have produced a single spectrum with a luminosity weighted
distance of 3 R$_e$ of sufficient S/N to determine an
age, metallicity and $\alpha$-element abundance ratio by comparison
with Lick line strength indices for Simple Stellar Population models.

We find evidence for slight rotation of amplitude 31$\pm$13 kms$^{-1}$ in 
the spheroid of NGC 3923 between 2 and 4 R$_e$ approximately 
along the major axis of the galaxy.

All 29 GCs examined in this study are consistent with being 
old ($\ga$10 Gyr) and none display any persuasive evidence for
being young enough to be associated with the shell forming event.
Our results therefore support the conclusions of \citet{sikkema06}
who, using HST ACS photometry for the GC system of 
NGC 3923, found no evidence of a young GC subpopulation.

The GCs examined display a wide range of metallicity from [Z/H] =
-1.8 to +0.35. In contrast to other studies 
\citep{Puzia05,Pierce3379,Pierce4649} we find no evidence for a 
significant subpopulation of GCs with 
extremely high ([$\alpha$/Fe]$>$0.5) $\alpha$-element abundance ratios. Our 
data also do not show any clear tendency for $\alpha$-element
abundance to increase with decreasing metallicity as had been found
in these studies. It must be noted
however that $\alpha$-element abundance ratios are not well constrained
at low metallicity and that the results of these previous studies may be
consistent within the errors with a universal value of [$\alpha$/Fe] = 0.3.
In agreement with both these studies and with measurements of the 
Milky Way GC population we find that the majority of our GCs with accurate
abundance ratios have [$\alpha$/Fe] consistent with 0.3.

The stellar population of the diffuse light of NGC 3923 between $\sim$
2 and 4R$_{e}$ is entirely consistent in age, metallicity and $\alpha$-element
abundance ratio with the red GC population.
It also agrees within the respective errors with the mean of the GC population
examined spectroscopically; however, there is a slight difference
in metallicity for the mean of the total GC population and the 
stellar population of the diffuse light of the galaxy. With the current sample size, 
it is impossible to determine if this offset is real or merely the result of 
a statistical fluctuation.

This similarity between the stellar populations of the spheroid of NGC 3923 at 
large radii and the redder GCs is in good agreement with previous findings, both
spectroscopic (in NGC 3115; \citet{NSK06}) and photometric (in NGC 4649; 
\citet{Forbes04}), and supports the view that the two populations formed coevally.

The fact that the stellar populations of the GCs studied here do 
not match those of the inner regions of NGC 3923 is most likely due to 
the fact that GC systems are observed to display variation in the
mix of red and blue populations with radii. Therefore the GC populations
which make up the region between 2-4R$_e$ are not likely to be the same
as those that inhabit the inner regions of the galaxy. Additionally 
GCs are thought to be vulnerable to disruption by tidal effects within the
inner regions of galaxies meaning that projection effects become a more
serious issue at smaller galactic radii, with GCs at large radii projected
onto the inner regions of the galaxy becoming more important.

By comparing our stellar population parameters for the spheroid of NGC 3923 
at large galactocentric radii with those of the inner regions 
of NGC 3923 measured by \cite{Thomas05} (using the index measurements 
of the inner R$_e$/10 by \cite{Beuing02}) and \cite{Denicolo05} (using index
measurements of the inner R$_e$/8), we find evidence for 
metallicity and age gradients in the spheroid of NGC 3923 but no evidence
of an [$\alpha$/Fe] abundance ratio gradient. Specifically \cite{Thomas05}
find a luminosity weighted age for the central region of the galaxy of 3.3 Gyr,
a metallicity of [Z/H] = 0.62 and [$\alpha$/Fe] = 0.31. 
This is in good agreement with the conclusions of \cite{Denicolo05}
who find an age of 2.6 Gyr, metallicity of [Z/H] $>$ 0.67 and a slightly lower
value of [$\alpha$/Fe] of around 0.14. This close agreement is to be expected 
as both studies make use of the same TMK04 models to derive their stellar 
population parameters.
We see therefore
that the outer regions of NGC 3923 are considerably older, less enriched in
metals, but have a similar $\alpha$-element abundance ratio as the 
inner regions of the galaxy. The first two points are not unexpected considering that the
values for \cite{Thomas05} and \cite{Denicolo05} are provided for apertures 
of $\leq$ R$_e$/8, and our
spectroscopy of the spheroid of NGC~3923 has a luminosity
weighted effective radius of 3R$_e$.

Such a situation is very similar to that found for the isolated elliptical 
NGC~821 by \cite{Proctor821} using longslit spectroscopy. In this case 
\cite{Proctor821} found that NGC~821 displayed a strong gradient in both metallicity 
and age, in the sense that metallicity decreases with radius and age increases with
radius. As in NGC~3923 the galaxy displays a constant value of [$\alpha$/Fe] 
close to 0.3. NGC~3923 
and NGC~821 display similar metallicity gradients of $\sim$-0.7 dex/$\Delta$logR
but the normalisation for NGC~3923 is shifted to higher metallicities
than in NGC~821. As discussed in the case of NGC~821 \citep{Proctor821}, this high value
of the metallicity gradient tends to rule out merger models for the formation
of NGC~3923, as merger models predict lower values for the metallicity
gradient due to the violent mixing that takes place during mergers tending 
to wash out gradients \citep{Kobayashi04}.

Another scenario considered by \cite{Proctor821} to explain the formation
of NGC~821 is where a minor merger event funnels gas to the inner region 
of the galaxy producing a minor star formation event and adding a layer 
of ``frosting" to the majority older population
already present. This star formation event presumably did not produce a
significant number of star clusters stable enough or at large enough galactocentric
radii for them to survive and be detectable at the present time. 
This scenario for the formation of NGC~821, can also be applied to the formation
of NGC~3923, where it can naturally explain the formation of
the shell structures. However the same problem that caused \cite{Proctor821}
to reject this explanation in the case of NGC~821 still applies to our
NGC~3923 data, namely that the gas forming the new stars would presumably
be of different metallicity and $\alpha$-element abundance than 
the galaxy at these radii-- it would then have to form stars over the correct timescale, 
with the appropriate enrichment such that they appeared to fall exactly on the 
trends of metallicity and $\alpha$-element abundance that the rest of the galaxy exhibits.
This appears to require an unlikely amount of fine tuning, however recent 
work by \citet{Serra07} shows that when examining a composite stellar population
using SSP models, it is the young component which dominates the
age determined for the composite. They further find that the chemical composition
(metallicity and [$\alpha$/Fe]) of the best fit SSP model closely follows that of the 
older population which dominates the mass. This explanation would also not rule out the
young age of around 1 Gyr determined by \citet{sikkema06} for the 
shell features of NGC~3923, as any 1 Gyr stellar population present in the 
inner regions could, with the correct choice of mass fraction, appear to be 
around 3 Gyr, when combined with a majority older stellar population.

A final possible formation scenario considered by \cite{Proctor821}
to explain the formation of NGC~821 is that the gas for the
recent starburst is fuelled by gas from the galaxy itself, perhaps through
gas cooling smoothly onto an inner disc. High spatial resolution IFU 
observations of the inner regions of NGC~3923 would help to search for
the signatures of such a structure. \citet{Serra07} note that if high enough 
S/N spectroscopy is available 
it is possible to disentangle the different stellar populations present
through examination of the discrepancy between the SSP parameters determined
using different Balmer-line indices. This result therefore provides a possible
route to determining the relevance of these models for the formation of NGC~3923.

\section{Conclusions}

We have obtained low-resolution spectra of 37 GC candidates
around the shell elliptical NGC 3923 using the GMOS on 
Gemini-South. By examination of the radial velocities of the
targets we find that 29 are GCs associated with the target galaxy.
Making use of a new technique of extracting integrated spectra at
extremely low surface brightness levels we have 
determined kinematic properties for the diffuse light of 
NGC 3923 out to around 200 arcsecs on both major and
minor axes,  corresponding to half light radii in the range 
2-4R$_e$. We find evidence for low amplitude rotation along the major
axis of around 30kms$^{-1}$. The detailed kinematic
properties of NGC 3923 (including its observed velocity
dispersion in this range) and its GC system will be studied
in more detail in a forthcoming paper.

We have measured the ages, metallicities and [$\alpha$/Fe]
ratios of the 29 GCs and compared them to those found found for the diffuse
light of NGC 3923 at the same projected radii. 
We find all of our GCs to have old ages 
($\geq$10 Gyr), with metallicities running from [Z/H] = -1.8 to +0.35 and 
$\alpha$-element abundance ratios generally consistent with a
constant value of [$\alpha$/Fe] = 0.3. We see no evidence in our data for the 
existence of a younger population of GCs associated with the 
$\sim$1 Gyr old merger event proposed to have produced the 
shell system of NGC 3923.
The diffuse light of NGC 3923 is observed to have properties
indistinguishable from those of the more metal rich (and redder)
 GCs.
 
 Our results provide support to the theory that the spheroids and GC 
 systems of galaxies are produced during the same star formation
 events, and that the study of the stellar populations of GC systems
 can provide important insights into the stellar populations of
 galaxies at galactocentric radii which at present are difficult to study
 using direct measurements of their integrated light.


\begin{table*} 
\begin{center}
\caption{Measured parameters for spectroscopically examined objects,
both confirmed GCs around NGC~3923 as well as fore and background contaminants.
In addition we provide the binned GC data and the measured indices for the
spheroid of NGC~3923.  
For the GCs: ID (and number of objects per bin for the binned data in parentheses), 
coordinates, {\it g} magnitude, {\it g--r} and {\it g--i} 
colours are from our GMOS imaging and are instrumental magnitudes
based on the standard Gemini/GMOS zeropoints.  
Heliocentric velocities are from the spectra presented in this work, as are
the measured line strength indices.
For the binned GC data the mean {\it g--i}  colours of the binned data
is provided, as is the error weighted mean line strength indices
for each bin, the errors for the line strengths are the 1$\sigma$ scatter
in the individual line strengths in each bin.
For the spheroid data the line strengths are provided as well as 
errors determined from a combination of the statistical uncertainty on 
the spectrum as well as the uncertainty in the velocity determination.
Data are presented here only for the main indices; the complete table of
measured indices is available with the electronic version of this article.
}
\begin{tabular}{lcccccccccc} 
\hline 
ID     & R.A.   & Dec.       &   {\it g$'$} & {\it g$'$--r$'$} & {\it g$'$--i$'$}  & V & H$\beta$ & Mg$b$ & Fe5270 & Fe5335\\ 
       & (J2000) & (J2000)   &  (mag) & (mag) &  (mag) & (km/s) & (\AA\,) & (\AA\,) & (\AA\,) & (\AA\,)\\
\hline
228 & 11:50:49.2 & -28:46:58.0 & 22.92 & 0.57 & 0.79 & 1587$\pm$66 & 3.86$\pm$0.29 & 0.68$\pm$0.31 & 1.76$\pm$0.34 & 0.51$\pm$0.40 \\
253 & 11:50:50.1 & -28:50:19.0 & 23.08 & 0.64 & 0.82 & 2203$\pm$74 & 1.98$\pm$0.32 & 1.64$\pm$0.35 & 0.23$\pm$0.39 & 0.84$\pm$0.42 \\
197 & 11:50:49.7 & -28:48:55.5 & 22.75 & 0.77 & 1.08 & 1727$\pm$29 & 1.87$\pm$0.25 & 3.87$\pm$0.25 & 2.17$\pm$0.27 & 1.56$\pm$0.30 \\
279 & 11:50:51.3 & -28:50:38.1 & 23.10 & 0.83 & 1.20 & 1807$\pm$41 & 1.40$\pm$0.32 & 4.00$\pm$0.33 & 2.77$\pm$0.35 & 3.03$\pm$0.36 \\
112 & 11:50:52.0 & -28:48:32.2 & 22.27 & 0.71 & 0.97 & 1763$\pm$26 & 2.26$\pm$0.18 & 2.59$\pm$0.19 & 1.80$\pm$0.21 & 1.57$\pm$0.23 \\
394 & 11:50:50.7 & -28:46:47.9 & 23.48 & 0.58 & 0.84 & 1851$\pm$69 & 1.71$\pm$0.47 & 0.79$\pm$0.49 & 1.95$\pm$0.52 & 0.55$\pm$0.61 \\
106 & 11:50:53.6 & -28:46:07.1 & 22.13 & 0.69 & 0.99 & 1733$\pm$26 & 1.96$\pm$0.16 & 1.96$\pm$0.17 & 1.36$\pm$0.18 & 1.50$\pm$0.21 \\
64   & 11:50:54.1 & -28:48:15.8 & 21.75 & 0.68 & 0.93 & 1770$\pm$27 & 2.05$\pm$0.13 & 1.50$\pm$0.13 & 1.27$\pm$0.15 & 1.17$\pm$0.17 \\
86   & 11:50:56.3 & -28:45:59.3 & 22.02 & 0.70 & 1.02 & 1784$\pm$22 & 2.13$\pm$0.15 & 2.47$\pm$0.15 & 1.57$\pm$0.17 & 2.00$\pm$0.18 \\
311 & 11:50:57.7 & -28:46:42.1 & 23.19 & 0.70 & 0.98 & 1666$\pm$43 & 2.16$\pm$0.35 & 2.25$\pm$0.39 & 1.68$\pm$0.40 & 2.80$\pm$0.44 \\
333 & 11:50:58.3 & -28:50:51.0 & 23.25 & 0.82 & 1.15 & 1925$\pm$55 & 2.78$\pm$0.40 & 3.97$\pm$0.42 & 3.10$\pm$0.44 & 2.90$\pm$0.48 \\
247 & 11:50:57.3 & -28:50:13.5 & 23.01 & 0.66 & 0.86 & 1829$\pm$48 & 2.36$\pm$0.31 & 2.02$\pm$0.33 & 1.57$\pm$0.36 & 0.99$\pm$0.43 \\
99   & 11:50:55.0 & -28:48:00.3 & 22.12 & 0.75 & 1.07 & 1654$\pm$22 & 1.42$\pm$0.17 & 2.91$\pm$0.18 & 2.17$\pm$0.19 & 2.36$\pm$0.20 \\
322 & 11:50:55.4 & -28:46:29.5 & 23.24 & 0.64 & 0.87 & 2099$\pm$50 & 1.86$\pm$0.40 & 1.17$\pm$0.41 & 1.11$\pm$0.44 & 1.08$\pm$0.49 \\
498 & 11:50:55.8 & -28:46:45.1 & 23.76 & 0.72 & 1.06 & 1600$\pm$49 & -0.06$\pm$0.60 & 2.52$\pm$0.62 & 3.49$\pm$0.62 & 4.02$\pm$0.65 \\
221 & 11:51:04.1 & -28:45:45.6 & 22.89 & 0.59 & 0.79 & 1805$\pm$53 & 2.11$\pm$0.29 & 1.41$\pm$0.30 & 0.87$\pm$0.33 & 1.38$\pm$0.38 \\
110 & 11:51:04.9 & -28:45:48.5 & 22.24 & 0.72 & 1.07 & 1550$\pm$24 & 1.45$\pm$0.18 & 2.90$\pm$0.18 & 2.14$\pm$0.19 & 1.65$\pm$0.22 \\
232 & 11:51:00.2 & -28:46:55.0 & 22.96 & 0.71 & 1.10 & 1808$\pm$30 & 1.49$\pm$0.31 & 2.71$\pm$0.31 & 2.17$\pm$0.34 & 1.55$\pm$0.38 \\
225 & 11:51:03.6 & -28:49:53.0 & 22.92 & 0.81 & 1.17 & 1926$\pm$37 & 1.57$\pm$0.28 & 4.22$\pm$0.29 & 1.66$\pm$0.32 & 2.45$\pm$0.34 \\
360 & 11:50:59.5 & -28:50:05.8 & 23.37 & 0.80 & 1.20 & 1476$\pm$39 & 1.61$\pm$0.40 & 3.37$\pm$0.48 & 2.19$\pm$0.46 & 3.60$\pm$0.48 \\
332 & 11:51:05.8 & -28:49:51.3 & 23.28 & 0.85 & 1.30 & 1639$\pm$41 & 0.54$\pm$0.43 & 4.52$\pm$0.40 & 3.67$\pm$0.41 & 3.67$\pm$0.43 \\
108 & 11:51:07.2 & -28:50:11.9 & 22.20 & 0.77 & 1.14 & 2046$\pm$32 & 1.41$\pm$0.18 & 3.52$\pm$0.18 & 1.88$\pm$0.19 & 2.32$\pm$0.21 \\
450 & 11:51:00.7 & -28:45:40.3 & 23.66 & 0.71 & 1.01 & 2114$\pm$39 & 3.02$\pm$0.54 & 2.79$\pm$0.56 & 1.80$\pm$0.58 & 1.88$\pm$0.65 \\
513 & 11:51:01.2 & -28:45:52.5 & 23.86 & 0.71 & 1.15 & 1671$\pm$61 & -1.10$\pm$0.73 & 3.86$\pm$0.63 & 1.53$\pm$0.71 & 1.81$\pm$0.76 \\
492 & 11:50:59.2 & -28:45:58.4 & 23.73 & 0.84 & 1.19 & 1746$\pm$50 & 1.99$\pm$0.53 & 3.43$\pm$0.54 & 2.13$\pm$0.58 & 1.45$\pm$0.64 \\
167 & 11:51:08.0 & -28:46:06.0 & 22.52 & 0.71 & 1.01 & 1806$\pm$22 & 1.65$\pm$0.22 & 2.74$\pm$0.22 & 1.41$\pm$0.24 & 0.88$\pm$0.28 \\
65   & 11:51:07.6 & -28:46:23.9 & 21.71 & 0.64 & 0.83 & 2295$\pm$30 & 1.77$\pm$0.13 & 1.20$\pm$0.14 & 0.70$\pm$0.15 & 1.09$\pm$0.17 \\
104 & 11:51:08.6 & -28:46:27.4 & 22.16 & 0.59 & 0.80 & 2152$\pm$45 & 2.41$\pm$0.17 & 1.05$\pm$0.18 & 0.94$\pm$0.20 & 0.85$\pm$0.22 \\
93   & 11:51:09.1 & -28:48:19.4 & 22.06 & 0.76 & 1.10 & 1754$\pm$25 & 1.93$\pm$0.16 & 3.82$\pm$0.16 & 0.98$\pm$0.18 & 2.01$\pm$0.20 \\
\hline
Bin 1 (8 GCs) &&& 22.81 & 0.61 &    0.82 ${^{0.90} _{0.78}}$  &&  2.20$\pm$0.49  & 1.26$\pm$0.38   &  1.05$\pm$0.42  &  0.89$\pm$0.26\\
Bin 2 (7 GCs) &&& 22.51 & 0.70 &    0.99 ${^{1.02} _{0.90}}$  &&  2.08$\pm$0.29  &  2.21$\pm$0.55  &  1.41$\pm$0.29  &  1.55$\pm$0.54\\
Bin 3 (7 GCs) &&& 22.58 & 0.74 &    1.09 ${^{1.14} _{1.02}}$  &&  1.67$\pm$0.33  &  3.33$\pm$0.36  &  1.72$\pm$0.43  &  2.05$\pm$0.38\\
Bin 4 (7 GCs) &&& 23.36 & 0.81 &    1.19 ${^{1.31} _{1.14}}$  &&  1.61$\pm$1.03  &  4.09$\pm$0.62  &  2.68$\pm$0.43  &  2.74$\pm$1.06\\
\hline
     All GCs (29) &&& 22.81 & 0.71 &  1.02  ${^{1.31} _{0.78}}$  &&  1.95$\pm$0.49  &  2.47$\pm$1.03  &  1.54$\pm$0.57  & 1.67$\pm$0.74\\
\hline
	NGC 3923 &&&&&&&  1.52$\pm$0.36  &  3.04$\pm$0.37  &  1.79$\pm$0.42  & 2.00$\pm$0.46\\
\hline
30 (Star)	& 11:50:54.5 & -28:50:19.5 & 21.17 & 0.76 & 1.03 & 91$\pm$34 &&&& \\
15 (Star)	& 11:50:56.9 & -28:50:03.9  & 20.15 & 0.95 & 1.31 & 19$\pm$37 &&&& \\
118 (Star)	& 11:51:02.7 & -28:45:54.1 & 22.28 &0.90 & 1.22 & 92$\pm$34 &&&& \\
40 (QSO)	& 11:51:06.3 & -28:46:46.6 & 21.45 & 0.50 & 0.46 & z$\sim$1.27 &&&& \\
56 (Star)	& 11:51:05.2 & -28:46:50.3 & 21.55 & 0.40 & 0.47 & 289$\pm$59 &&&& \\
383 (Galaxy)	& 11:51:04.5 & -28:50:38.2 & 23.47 & 0.90 & 1.14 & z$\sim$0.36 &&&& \\
113 (Galaxy)	& 11:51:10.7 & -28:45:38.8 & 22.22 & 0.83 & 1.21 & z$\sim$0.29 &&&& \\
17 (Star)	& 11:51:09.9 & -28:49:42.3 & 20.21 & 0.54 & 0.69 & 111$\pm$62 &&&& \\
	
\label{tabobs}
\end{tabular} 
\end{center} 
\end{table*}


\begin{table*} 
\begin{center}
\caption{Derived stellar population parameters for the globular clusters and spheroid
of NGC 3923. Age, [Fe/H], [$\alpha$/Fe] and [Z/H] derived using the $\chi^2$ minimization
method described in Section 4.2. }
\begin{tabular}{lcccccc} 
\hline 
ID    & 	Age 	& 	[Fe/H] 	& 	[$\alpha$/Fe] 	&	[Z/H]	   &	S/N	& Comments  \\ 
        &	(Gyr) &	(dex)	&	(dex)		&	(dex)  &     \\
\hline
228	& 10.0 ${^{+3.0} _{-2.3}}$	&	-2.25$\pm$0.37	& 	0.48$\pm$0.40 & -1.80$\pm$0.17 & 24.4 & \\
253	& 15.0 ${^{+8.8} _{-5.5}}$	&	-2.28$\pm$0.34	& 	0.80$\pm$0.30 & -1.53$\pm$0.29 & 22.5 & \\
197	& 6.0	 ${^{+2.7} _{-1.8}}$	&	-0.49$\pm$0.17	& 	0.44$\pm$0.10 & -0.08$\pm$0.13 & 29.4 & \\
279	& 15.0 ${^{+9.9} _{-5.9}}$	&	-0.57$\pm$0.19	& 	0.42$\pm$0.13 & -0.18$\pm$0.16 & 22.2 & \\
112	& 7.9	 ${^{+3.4} _{-2.4}}$	&	-0.88$\pm$0.13	&	0.30$\pm$ 0.08 & -0.60$\pm$0.15 & 39.4 & \\
394	& 11.9 ${^{+4.9} _{-3.5}}$	&	-1.31$\pm$0.27	&	0.38$\pm$0.41 & -0.95$\pm$0.28 & 15.6 & Fit unstable. \\
106	& 11.9 ${^{+1.9} _{-1.7}}$	&	-1.19$\pm$0.12	&	0.15$\pm$0.12 & -1.05$\pm$0.08 & 44.7 & \\
64	& 11.9 ${^{+2.0} _{-1.7}}$	&	-1.37$\pm$0.10	&	0.21$\pm$0.10 & -1.18$\pm$0.06 & 56.1 & \\
86	& 15.0 ${^{+6.3} _{-4.4}}$	&	-1.15$\pm$0.15	&	0.34$\pm$0.11 & -0.83$\pm$0.12 & 47.7 & \\
311	& 6.7 ${^{+7.5} _{-3.5}}$		&	-0.72$\pm$0.37	&	0.18$\pm$0.22 & -0.55$\pm$0.27 & 20.0 & \\
333	& 15.0 ${^{+28.0} _{-9.8}}$	&	-0.27$\pm$0.26	&	0.21$\pm$0.16 & -0.08$\pm$0.30 &17.8 & No Balmer lines in fit.\\
247	& 14.1 ${^{+8.6} _{-5.3}}$	&	-1.35$\pm$0.28	&	0.32$\pm$0.24 & -1.05$\pm$0.21 & 22.9 & \\
99	& 12.6 ${^{+2.7} _{-2.2}}$	&	-0.57$\pm$0.11	&	0.15$\pm$0.10 & -0.43$\pm$0.08 & 43.4 & \\
322	& 11.2 ${^{+4.5} _{-3.2}}$	&	-1.25$\pm$0.26	&	0.24$\pm$0.30 & -1.03$\pm$0.20 & 19.3 & \\
498	& 15.0 ${^{+15.0} _{-7.5}}$	&	0.13	$\pm$0.25	&	-0.30$\pm$0.12 & -0.15$\pm$0.29 & 12.4 & Extremely poor fit \\
221	& 10.6 ${^{+2.7} _{-2.1}}$	&	-1.45$\pm$0.25	&	0.24$\pm$0.26 & -1.23$\pm$0.11 & 24.9 & \\
110	& 15.0 ${^{+4.7} _{-3.6}}$	&	-0.76$\pm$0.15	&	0.30$\pm$0.10 & -0.48$\pm$0.11 & 40.6 & \\
232	& 11.9 ${^{+4.2} _{-3.1}}$	&	-0.78$\pm$0.18	&	0.24$\pm$0.15 & -0.55$\pm$0.16 & 22.9 & \\
225	& 11.9 ${^{+4.8} _{-3.4}}$	&	-0.85$\pm$0.24	&	0.32$\pm$0.36 & -0.55$\pm$0.24 & 25.7 & \\
360	& 10.6 ${^{+5.9} _{-3.8}}$	&	-0.77$\pm$0.24	&	0.53$\pm$0.20 & -0.28$\pm$0.14 & 17.0 & \\
332	& 15.0 ${^{+8.6} _{-5.5}}$	&	0.15$\pm$0.16		&	0.03	$\pm$0.12 & 0.18$\pm$0.18 &18.7 & Fit unstable. \\
108	& 14.1 ${^{+2.2} _{-1.9}}$	&	-0.45$\pm$0.07	&	0.24$\pm$0.05 & -0.23$\pm$0.05 & 41.9 & \\
450	& 10.0 ${^{+6.2} _{-3.8}}$	&	-0.61$\pm$0.28	&	0.38$\pm$0.19 & -0.25$\pm$0.22 & 13.2 & \\
513	& 13.3 ${^{+12.0} _{-6.3}}$	&	-0.58$\pm$0.28	&	0.40$\pm$0.22 & -0.20$\pm$0.23 & 11.4 & \\
492	& 8.9 ${^{+7.9} _{-4.2}}$		&	-0.23$\pm$0.24	&	0.62$\pm$0.18 & 0.35$\pm$0.25 & 13.0 & \\
167	& 12.6 ${^{+3.7} _{-2.9}}$	&	-1.30$\pm$0.15	&	0.50$\pm$0.13 & -0.83$\pm$0.12 & 33.3 & \\
65	& 11.9 ${^{+2.1} _{-1.8}}$	&	-1.60$\pm$0.15	&	0.27$\pm$0.15 & -1.35$\pm$0.07 & 53.5 & \\
104	& 11.9 ${^{+1.8} _{-1.6}}$	&	-1.73$\pm$0.17	&	0.38$\pm$0.17 & -1.38$\pm$0.07 & 41.8 & \\
93	& 15.0 ${^{+6.2} _{-4.4}}$	&	-0.75$\pm$0.14	&	0.18$\pm$0.17 & -0.58$\pm$0.19 & 45.6 & \\
\hline
Bin 1 (8 GCs) & 11.9 ${^{+2.9} _{-2.3}}$	&	-1.60$\pm$0.29	&	0.24$\pm$0.30 & -1.38$\pm$0.12 & & \\
Bin 2 (7 GCs) & 12.6 ${^{+3.6} _{-2.8}}$	&	-1.18$\pm$0.15		&	0.24$\pm$0.16 & -0.95$\pm$0.10 & & \\
Bin 3 (7 GCs) & 15.0 ${^{+5.6} _{-4.1}}$	&	-0.70$\pm$0.16		&	0.27$\pm$0.12 & -0.45$\pm$0.11 & & \\
Bin 4 (7 GCs) & 13.3 ${^{+7.8} _{-4.9}}$	&	-0.35$\pm$0.20		&	0.24$\pm$0.16 & -0.13$\pm$0.15 & & \\
\hline
All GCs  (29 GCs) & 15.0 ${^{+15.4} _{-7.6}}$	&	-1.00$\pm$0.37		&	0.21$\pm$0.29 & -0.80$\pm$0.31& & \\
\hline
NGC 3923 & 15.0 ${^{+5.5} _{-4.0}}$	&	-0.65$\pm$0.17		&	0.34$\pm$0.13 & -0.33$\pm$0.14 & & \\
\hline
\label{tabres}
\end{tabular} 
\end{center} 
\end{table*}

\begin{table} 
\begin{center}
\caption{Measured velocities for GCs and NGC~3923 spheroid. ID, RA, DEC,
target velocity (including fore and background objects) and velocity of NGC~3923
spheroid measured in the same slitlet.}
\begin{tabular}{lcccc} 
\hline 
ID    & 	R.A. & DEC. &   V$_{target}$ & V$_{NGC~3923}$  \\
& (J2000) & (J2000)   & (km/s) & (km/s) \\
\hline
228  &    11:50:49.2  &   -28:46:58.0   &   1587$\pm$66  &    1866$\pm$72 \\
253  &    11:50:50.1  &   -28:50:19.0  &    2203$\pm$74  &    1812$\pm$64 \\
197  &    11:50:49.7  &   -28:48:55.5  &    1727$\pm$29 &     1859$\pm$37 \\
279  &    11:50:51.3  &   -28:50:38.1  &    1807$\pm$41   &   1772$\pm$101 \\
112  &    11:50:52.0 &    -28:48:32.2 &     1763$\pm$26   &   1815$\pm$34 \\
394  &    11:50:50.7 &    -28:46:47.9 &     1851$\pm$69   &   1843$\pm$64 \\
106  &    11:50:53.6  &   -28:46:07.1 &     1733$\pm$26   &   1972$\pm$106 \\
30    &   11:50:54.5  &   -28:50:19.5   &   91$\pm$34   &   1750$\pm$90 \\
64    &   11:50:54.1  &   -28:48:15.8  &    1770$\pm$27  &    1855$\pm$21 \\
86    &   11:50:56.3   &  -28:45:59.3   &   1784$\pm$22  &    1830$\pm$50 \\
311  &    11:50:57.7 &    -28:46:42.1 &     1666$\pm$43  &    1879$\pm$41 \\
333  &    11:50:58.3  &   -28:50:51.0 &     1925$\pm$55  &    1952$\pm$48 \\
15    &   11:50:56.9   &  -28:50:03.9   &   19$\pm$37  &    1917$\pm$33 \\
247  &    11:50:57.3 &    -28:50:13.5  &    1829$\pm$48  &    1874$\pm$23 \\
99    &   11:50:55.0  &   -28:48:00.3  &    1654$\pm$22  &    1874$\pm$32 \\
322  &    11:50:55.4  &   -28:46:29.5 &     2099$\pm$50  &    1817$\pm$58 \\
498  &    11:50:55.8 &    -28:46:45.1 &     1600$\pm$49  &    1867$\pm$32 \\
221  &    11:51:04.1  &   -28:45:45.6  &    1805$\pm$53  &    1972$\pm$45 \\
110  &    11:51:04.9  &   -28:45:48.5  &    1550$\pm$24  &    1910$\pm$54 \\
118  &    11:51:02.7 &    -28:45:54.1   &   92$\pm$34   &   1909$\pm$45 \\
232  &    11:51:00.2 &    -28:46:55.0   &   1808$\pm$30  &    1853$\pm$27 \\
225  &    11:51:03.6 &    -28:49:53.0   &   1926$\pm$37  &    1878$\pm$29 \\
360  &    11:50:59.5  &   -28:50:05.8  &    1476$\pm$39  &    1848$\pm$22 \\
332  &    11:51:05.8  &   -28:49:51.3  &    1639$\pm$41  &    1881$\pm$37 \\
108  &    11:51:07.2  &   -28:50:11.9 &     2046$\pm$32  &    1984$\pm$84 \\
40    &   11:51:06.3  &   -28:46:46.6  &    z$\sim$1.27  &    1843$\pm$24 \\
56    &   11:51:05.2  &   -28:46:50.3   &   289$\pm$59   &   1858$\pm$25 \\
450 &     11:51:00.7 &    -28:45:40.3  &    2114$\pm$39  &    1923$\pm$75 \\
513  &    11:51:01.2 &    -28:45:52.5   &   1671$\pm$61   &   1913$\pm$49 \\
492  &    11:50:59.8  &   -28:45:58.4   &   1746$\pm$50  &    1911$\pm$59 \\
383  &    11:51:04.5  &   -28:50:38.2   &   z$\sim$0.36  &    1822$\pm$45 \\
167  &    11:51:08.0  &   -28:46:06.0   &   1806$\pm$22  &    1907$\pm$42 \\
65    &   11:51:07.6   &  -28:46:23.9   &   2295$\pm$30   &   1860$\pm$38 \\
104 &     11:51:08.6  &   -28:46:27.4   &   2152$\pm$45   &   1900$\pm$42 \\
113  &    11:51:10.7  &   -28:45:38.8   &   z$\sim$0.29   &   1872$\pm$91 \\
93    &   11:51:09.1  &   -28:48:19.4   &   1754$\pm$25   &   1878$\pm$29 \\
17    &   11:51:09.9 &    -28:49:42.8   &   111$\pm$62   &   1929$\pm$51 \\
\hline
\label{tabkin}
\end{tabular} 
\end{center} 
\end{table}


\section{Acknowledgements}

The authors would like to acknowledge Russell Smith 
for input and discussions which greatly improved this work
and Bryan Miller for making his GMOS-S observations of
Lick standard stars available to us. We also thank the anonymous
referee for several suggestions which improved this work.
MAN acknowledges financial support from the STFC. SEZ 
acknowledges support from NSF award AST-0406891.
DF acknowledges support from the ARC.

Based on observations obtained at the Gemini Observatory, 
which is operated by the Association of Universities for Research 
in Astronomy, Inc., under a cooperative agreement with the NSF 
on behalf of the Gemini partnership: the National Science 
Foundation (United States), the Science and Technology Facilities 
Council (United Kingdom), the National Research Council (Canada), 
CONICYT (Chile), the Australian Research Council (Australia), 
CNPq (Brazil) and CONICET (Argentina).

\bibliographystyle{mn2e}
\bibliography{references}
\label{lastpage}


\end{document}